\documentclass[12pt]{article}
\usepackage{putex}
\usepackage{graphicx}
\usepackage{amsmath,amssymb}

\newcommand{\zh}{\ensuremath{z_0}}

\begin{document}

\preprint{PUPT-2258 \\ LMU-ASC 10/08}

\institution{PU}{Joseph Henry Laboratories, Princeton University, Princeton, NJ 08544, USA}
\institution{MaxPlanck}{Ludwig-Maximilians-Universit\"at, Department f\"ur Physik, Theresienstrasse 37, \cr 80333 M\"unchen, Germany}

\title{Linearized hydrodynamics from probe-sources in the gauge-string duality}

\authors{Steven S. Gubser\worksat{\PU,}\footnote{e-mail: {\tt ssgubser@Princeton.EDU}} and Amos Yarom\worksat{\MaxPlanck,}\footnote{e-mail: {\tt yarom@theorie.physik.uni-muenchen.de}}}

\abstract{We study the response of an infinite, asymptotically static $\mathcal{N}=4$ plasma to a generic localized source in the probe approximation.  At large distances, the energy momentum tensor of the plasma includes a term which satisfies the constitutive relations of linearized hydrodynamics, but it can also include a non-hydrodynamical term which contributes at the same order as viscous corrections, or even at leading order in some cases.
The conditions for the appearance of a laminar wake far behind the source and its relevance for phenomenological models used to explain di-hadron correlations are discussed.  We also consider the energy momentum tensor near the source, where the hydrodynamical approximation can be expected to break down. Our analysis encompasses a wide range of sources which are localized in the bulk of AdS, including trailing strings, mesonic and baryonic configurations of strings, and point particles.}

\date{March 2008}

\maketitle

\tableofcontents

\section{Motivation}
\label{S:Introduction}

As a parton moves through a quark-gluon plasma, it loses energy to the medium.  It is interesting to ask where this energy goes.  At sufficiently large length scales, the main channels available are hydrodynamical.  There are two different hydrodynamical modes.  One is sound. Assuming the parton is moving at a speed $v$ greater than the speed of sound $c_s$, the signature of energy loss into the sound mode is a sonic boom.
The other mode is dispersive, having to do with the formation of laminar wakes (or diffusion wakes---we will not attempt to draw a distinction between them).
Neglecting effects related to the expansion or other collective flows of the medium, the laminar wake  is a stream of fluid behind the parton, moving in the same direction as the parton.
Once the plasma hadronizes, we can expect that a sonic boom will lead to enhanced particle production at the Mach angle $\theta = \cos^{-1} c_s/v$, while a hadronized wake will lead to intensified particle production in the direction of the parton's motion \cite{Stoecker:2004qu,Casalderrey-Solana:2004qm}.

Evidence for such effects may be obtained from histograms of the azimuthal angle between pairs of energetic hadrons
produced in heavy ion collisions \cite{Adler:2005ee,Adams:2005ph}. In \cite{Adler:2005ee} it was observed that for certain momenta the two-point correlation function between jets emitted from the plasma is peaked at an azimuthal angle of roughly $\pi\pm1.2$ radians and has a minimum at $\pi$ radians. This is described as ``jet-splitting''. Jet-splitting is suggestive of a sonic boom: a standard interpretation is that one energetic hadron (the ``trigger'' or ``near-side'' hadron) came from a hard parton that exited the plasma without losing much energy, and the other one (the ``associated'' or ``away-side'' hadron) was produced from the sonic boom caused by another hard parton whose momentum was opposite the first, at least in azimuthal angle.
In \cite{Adams:2005ph}, with more inclusive momentum cuts (and also the greater rapidity acceptance characteristic of STAR),
instead of jet-splitting, a broad peak was found, centered around $\pi$.\footnote{A recent analysis \cite{Ulery:2008pj} of STAR data using three-point functions exhibits a triple peak structure, where a central peak at $\Delta\phi_1=\Delta\phi_2=\pi$ is accompanied by two peaks of equal heights at $\Delta\phi_1\approx \pi \pm 1.4$ and $\Delta\phi_2\approx \pi \mp 1.4$. These latter two peaks are consistent with a sonic boom, while the central peak may be evidence for a diffusion wake.}

It is challenging to make an unambiguous connection between these results and hydrodynamics, but there are several notable efforts.
For example, in \cite{Casalderrey-Solana:2004qm} it was shown, using a hydrodynamical model and Cooper-Frye hadronization, that jet-splitting doesn't occur unless the diffusion wake is suppressed relative to the sonic boom; and in \cite{Renk:2006mv}, jet-splitting, in approximate agreement with PHENIX di-hadron correlators, was predicted using a model where three quarters of the energy goes into the sonic boom.  The discussion so far is anything but an exhaustive account of either the experimental or the theoretical literature on medium-induced modifications of jet structure.  Recent brief discussions can be found in \cite{CasalderreySolana:2007km,Zajc:2008ey}, while a broader treatment with more extensive references is included in \cite{Abreu:2007kv}.

In both \cite{Casalderrey-Solana:2004qm} and in \cite{Renk:2006mv} the authors have tuned the relative amount of energy going into sound modes and diffusion modes by hand.  Indeed, it is challenging to predict from QCD the relative strength of the sonic boom and the diffusion wake produced by a hard parton---assuming, of course, that sonic booms and diffusion wakes are the right language for describing the energy loss at length scales significantly larger than $1\,{\rm fm}$.  The essential difficulty is in connecting the short-distance physics, which is perturbative (or at least partly perturbative) and the long-distance hydrodynamical regime.  It might therefore be enlightening to consider similar phenomena in ${\cal N}=4$ super-Yang-Mills theory, where the methods of the gauge-string duality \cite{Maldacena:1997re,Gubser:1998bc,Witten:1998qj} allow detailed calculations which become reliable in the large $N$, large $g_{YM}^2 N$ limit.  Indeed, a number of papers \cite{Friess:2006fk, Gubser:2007nd, Gubser:2007xz, Chesler:2007sv, Gubser:2007ni, Gubser:2007ga, Chesler:2007an} are devoted to studying the sonic boom and diffusion wake produced by the trailing string of \cite{Herzog:2006gh,Gubser:2006bz}, which represents a heavy quark moving at constant velocity through a thermal medium of ${\cal N}=4$ gauge theory.  A strong diffusion wake is predicted in these works, similar to ``scenario 1'' of \cite{Casalderrey-Solana:2004qm}, which led to no jet-splitting after Cooper-Frye hadronization.  On the other hand, in \cite{Gubser:2007zr}, we showed that a string configuration representing a heavy-quark meson as described in \cite{Liu:2006nn,Chernicoff:2006hi} produces a sonic boom but no diffusion wake.

The aim of this paper is to get a stronger foothold on the hydrodynamic behavior of the ${\cal N}=4$ plasma due to a generic probe source, with an eye toward the phenomenologically interesting question of when a diffusion wake arises and how strong it is relative to the sonic boom.  Our main result, which characterizes the linear response of the stress-tensor of the boundary theory to the source and provides a comparison to linear hydrodynamics, is presented in section~\ref{S:Summary}. Hydrodynamic modes in the strongly coupled $\mathcal{N}=4$ plasma have been widely studied following \cite{Policastro:2001yc}, and recent works include \cite{Bhattacharyya:2007vs, Heller:2007qt, Gubser:2007ga, Chesler:2007sv, Baier:2007ix, Bhattacharyya:2007jc}, where more extensive references may be found.
In section \ref{S:EMTensor} and \ref{S:Long} we give details of our computation. In section \ref{S:Short} we discuss the near field of the source, which seems non-hydrodynamical in nature \cite{Noronha:2007xe}. In section \ref{S:Examples} we apply our methods to strings and point particle sources.

\section{Summary of results}
\label{S:Summary}

We start with an action
 \eqn{GenericAction}{
  S = {1 \over 2\kappa_5^2}
    \int d^5 x \, \sqrt{-G} \left[ R + {12 \over L^2} \right] +
    S_M \,,
 }
where $S_M$ describes a localized source. Without the source, one finds the standard translationally invariant $AdS_5$-Schwarzschild solution to the Einstein equations
\begin{equation}
\label{E:AdSBH}
	ds^2 = \alpha^2 \left(-h dt^2+\sum_{i=1}^3 dx_i^2 +
	  {dz^2 \over h} \right)
\end{equation}
where
\begin{equation}
\label{E:alphaandh}
	\alpha(z) = {L \over z} \,,\qquad
	h(z) = 1 - \left(\frac{z}{\zh}\right)^4 \,.
\end{equation}
The asymptotically AdS boundary is located at $z=0$ and the black hole horizon is at $z=z_0$.
The AdS/CFT duality relates strings moving in an asymptotically AdS${}_5\times$S${}^5$ background to $\mathcal{N}=4$ Super Yang Mills theory. We have omitted the $S^5$ part of the metric, assuming that the equations of motion of the source and the Einstein equations can be consistently truncated to the AdS${}_5$ directions. In \cite{Herzog:2007kh} the role of the motion of a string-source along the $S^5$ directions was studied in a rotating black hole background.

Treating a source in the probe approximation means determining its motion by extremizing $S_M$ in the background \eno{E:AdSBH} and then solving the linearized Einstein equations in the presence of the source to find a perturbed metric.  Let the energy momentum tensor of the source be denoted $J_{\mu\nu}(t,\vec{x},z)$, where indices $\mu$ and $\nu$ run over the five dimensions of $AdS_5$-Schwarzschild.  The tensor $J_{\mu\nu}$ is defined by the first variation of $S_M$ with respect to the metric:
 \eqn{SMvary}{
  \delta S_M = {1 \over 2L^3}
    \int d^5 x \, \sqrt{-G} \, \delta G^{\mu\nu} J_{\mu\nu} \,,
 }
and the Einstein equations (before linearization) read
 \eqn{EinsteinEqs}{
  R_{\mu\nu} - {1 \over 2} R G_{\mu\nu} - {6 \over L^2} G_{\mu\nu}
    = {\kappa_5^2 \over L^3} J_{\mu\nu} \,.
 }
The factors of $1/L^3$ in front of $J_{\mu\nu}$ in \eno{SMvary} and~\eno{EinsteinEqs} are included for later convenience.

The gauge-string duality relates the metric perturbations near the boundary of $AdS_5$ to the expectation value of the energy-momentum tensor of the boundary gauge theory, which we denote by $\langle T_{mn} \rangle$.  The indices $m$ and $n$ run from $0$ to $3$, and $5$ represents the radial AdS direction: that is, $x^5 = z$.  Our first main result is a generalization of the observation, first made quantitatively in \cite{Friess:2006fk}, that $T_{mn}$ is not necessarily conserved in the presence of a source: instead,
\begin{equation}
\label{E:Tmnsourced}
    \partial^m \langle T_{mn}(t,\vec{x}) \rangle =
      J_{n5}^{(3)}(t,\vec{x}) \,,
\end{equation}
where $J_{\mu\nu}^{(3)}(t,\vec{x})$ is defined through
 \eqn{Jseries}{
  J_{\mu\nu}(t,\vec{x},z) =
    \sum_{a=a_0}^\infty J^{(a)}_{\mu\nu}(t,\vec{x}) z^a \,,
 }
where $a$ is integer-valued and $a_0 \geq -1$. There are some caveats to \eno{E:Tmnsourced}: if $J_{55}^{(2)} \neq 0$ then $\langle T_{mn} \rangle$ will have a trace anomaly. In addition,  $\langle T_{mn} \rangle$ can have divergent terms which can contribute to the non-conservation law \eno{E:Tmnsourced}. These divergences can usually be ignored in phenomenological applications because their contribution to $\langle T_{mn} \rangle$ is confined to the location of the probe. In section~\ref{S:EMTensor} and appendix~\ref{A:HoloRG} we derive the result \eno{E:Tmnsourced} and explain this caveat in more detail.

Perhaps the most phenomenologically interesting information about the response of the medium to the source comes from the long distance asymptotics, where $\langle T_{mn} \rangle$ is expected to approximately satisfy the constitutive relations of hydrodynamics.\footnote{
An interesting recent study \cite{Noronha:2007xe} explores the degree to which the non-linear constitutive relations are maintained in the near-field regime of the trailing string, building upon results of \cite{Yarom:2007ap,Gubser:2007nd,Yarom:2007ni}.  Comparisons were also made for the trailing string between string theory results and linearized hydrodynamics in \cite{Chesler:2007sv}, and between string theory and a regulated version of linearized hydro results in \cite{Gubser:2007xz}.}
Our second main result characterizes the large distance behavior of $\langle T_{mn} \rangle$ in Fourier space,
\begin{equation}
\label{E:TmnFourier}
 	\langle T_{mn}(t,\vec{x})\rangle = \int \frac{d\omega}{2\pi}\frac{d^3k}{(2\pi)^3}e^{-i\omega t+i\vec{k}\cdot\vec{x}}\langle T_{mn}(\omega,\vec{k})\rangle\,.
\end{equation}
Still within the probe approximation, we find that the small $k$ behavior of the boundary theory stress-energy tensor $\langle T_{mn} \rangle$ is given by
\begin{equation}
\label{MainResult}
    \langle T_{mn} \rangle = \langle T_{mn} \rangle_{\rm bath}+T_{mn}^{\rm hydro}
        +\mathcal{F}_{mn}+\hbox{(corrections)}
\end{equation}
where
\begin{equation}
 	\langle T_{mn} \rangle_{\rm bath} = \frac{\pi^2}{8}(N^2-1) T^4 \,\hbox{diag}\left(3,1,1,1\right)
\end{equation}
is the energy momentum of the thermal bath, $N$ is the number of colors in the ${\cal N}=4$ gauge theory, and $T$ is the temperature.
The correction terms, labeled $\hbox{(corrections)}$ in \eno{MainResult}, are smaller than $T_{mn}^{\rm hydro}$ by a factor of at least ${\cal O}(k)$. If a multipole expansion of the drag force density, $J_{n5}^{(3)}$, includes a monopole term then these corrections are smaller than $T_{mn}^{\rm hydro}$ by a factor of ${\cal O}(k^2)$.
The tensor ${\cal F}_{mn}$ appearing in the third term of \eno{MainResult} is defined as follows:
\begin{equation}\label{FResult}
    \mathcal{F}_{ij} = {\rm F}\left[\int_z^{z_0}
      {d\zeta \over \zeta^3} J_{(ij)}(\zeta) \right]\qquad
	\mathcal{F}_{m0}=\mathcal{F}_{0m}=0 \,,
\end{equation}
where ${\rm F}\left[f(z)\right]$ means the finite part of $f(z)$ when taking the $z \to 0$ limit, and the index $i$ runs from 1 to 3.  The notation $(ij)$ means to take the symmetric traceless part of a $3 \times 3$ matrix:
 \eqn{TracelessPart}{
  X_{(ij)} = {1 \over 2} (X_{ij} + X_{ji}) - {1 \over 3} \delta_{ij}
    \delta^{gh} X_{gh} \,.
 }
As in \eqref{E:Tmnsourced}, we have discarded from \eno{MainResult} divergent contributions to the stress-energy tensor, a trace anomaly term and additional finite terms which are related to logarithmic divergences.
We discuss these terms in detail in sections \ref{S:EMTensor}, \ref{S:Long} and appendix \ref{A:HoloRG} where we holographically renormalize the stress-energy tensor.

The tensor $T_{mn}^{\rm hydro}$ appearing in the second term of \eno{MainResult} is defined by two requirements.  First, it should satisfy the constitutive relations of linearized hydrodynamics, which means that it can be written in the form
\begin{equation}\label{THydro}
	T^{mn}_{\rm hydro} = \begin{pmatrix}
				\epsilon & & S_i & \\
				& & &\\
				S_i & &\frac{1}{3}\epsilon \delta_{ij} - \frac{1}{2\pi T} \partial_{(i}S_{j)} & \\
				& & &
	                     \end{pmatrix} \,,
\end{equation}
where we have again used the notation \eno{TracelessPart}.  Note that $\epsilon$ is the deviation of $\langle T^{00} \rangle$ from its value in the thermal bath, whereas $S_i$ is the complete Poynting vector, simply because neither the bath nor ${\cal F}_{mn}$ contributes to it.  
The second requirement on $T_{mn}^{\rm hydro}$ is that it should be traceless and should obey the conservation equation that follows from plugging \eno{MainResult} into \eno{E:Tmnsourced}: that is,
 \eqn{TmnHydroSourced}{
  \partial^m T_{mn}^{\rm hydro} = F_n^{\rm hydro} \equiv
    J_{n5}^{(3)} - \partial^m {\cal F}_{mn} \,.
 }
The explicit values of $\epsilon$ and $S$ after implementing conditions \eqref{THydro} and \eqref{TmnHydroSourced} can be found in \eqref{E:resummed}.
We could summarize \eno{MainResult} by saying that $T_{mn}^{\rm hydro}$ captures most of the leading asymptotics at small $k$, and ${\cal F}_{mn}$ characterizes the deviations from the leading hydrodynamical behavior. Deviations of the quark gluon plasma from hydrodynamical behavior was also observed in \cite{Bhattacharyya:2007vs} when the gauge theory was placed on a sphere. We note that the deviations we observe here are different in nature from those in \cite{Bhattacharyya:2007vs} since they are associated with excitations created by the probe source and not with inherent properties of the fluid. In \cite{Gubser:2007ni,Gubser:2007zr,Chesler:2007sv} these deviations have been studied for the specific case of a heavy quark and heavy meson source.
We emphasize that \eqref{THydro} has been obtained by solving the equations of motion for the metric perturbations in a black hole background---we have not assumed its form a priori. Thus, the constitutive relations we've obtained in \eqref{THydro} provide an alternate derivation of the standard result $\eta/s=1/4\pi$ \cite{Policastro:2001yc}.

Obtaining our two main results, \eno{E:Tmnsourced} and \eno{MainResult}, requires some technical assumptions.
To obtain \eno{E:Tmnsourced}, we must assume that \eqref{Jseries}, a power series expansion of $J_{\mu\nu}$ near the boundary, exists.
To obtain \eno{MainResult}, we must additionally assume that a near boundary power series expansion exists also for the Fourier coefficients of $J_{\mu\nu}(\omega,\vec{k},z)$ 
defined through
 \eqn{FourierTransform}{
  J_{\mu\nu}(t,\vec{x},z) =
    \int {d\omega \over 2\pi} {d^3 k \over (2\pi)^3} \,
      e^{-i\omega t + i \vec{k} \cdot \vec{x}}
     J_{\mu\nu}(\omega,\vec{k},z) \,,
 }
and that the source is in some sense localized in the $x^m$ directions.  The latter assumption may be made precise in one of three ways:
 \begin{enumerate}
  \item Either $J_{\mu\nu}(\omega,\vec{k},z)$ should be analytic in $\omega$ and $\vec{k}$;
  \item Or it should be expressible in the form
 \eqn{ExpressJ}{
  J_{\mu\nu}(\omega,\vec{k},z) = \Delta(\omega,\vec{k})
    j_{\mu\nu}(\omega,\vec{k},z) \,,
 }
where $j_{\mu\nu}(\omega,\vec{k},z)$ is analytic in $\omega$ and $\vec{k}$ but $\Delta(\omega,\vec{k})$ need not be;
  \item Or it should be expressible as a sum of terms of the form shown on the right hand side of~\eno{ExpressJ}.
 \end{enumerate}
In all three cases one must require that $J_{\mu\nu}(t,\vec{x},z)$ is conserved: that is, $D^\mu J_{\mu\nu} = 0$ with $D_{\mu}$ the covariant derivative in the background \eqref{E:AdSBH}.
We must also make some mild assumptions about the behavior of $J_{\mu\nu}$ near the horizon to ensure that a non-singular metric response there is possible.  The precise form of these assumptions will be discussed in section~\ref{SS:Source} and \ref{S:Long}.

\section{The boundary stress tensor and its non-conservation}
\label{S:EMTensor}

Our overall goal is to evaluate the energy momentum tensor of the boundary theory $\langle T_{mn} \rangle$ in terms of the energy momentum tensor $J_{\mu\nu}$ of a probe-source moving in the $AdS_5$-Schwarzschild background.  The focus of this section is to extract the maximum possible information from a near-boundary analysis and to put it in a form that will make it easy to consider the large distance asymptotics, which we do in section~\ref{S:Long}.  In section~\ref{SS:Source} we discuss the consequences of the conservation equation for the energy momentum tensor $J_{\mu\nu}$ of the probe.  In section~\ref{SS:MetricPerturbations} we perform a near-boundary analysis of the perturbations of the five-dimensional metric and explain how $\langle T_{mn} \rangle$ can be extracted from them.  In section~\ref{SS:BoundaryStress} we use constraints on $\langle T_{mn} \rangle$ to give a particular parametrization of it in terms of five undetermined coefficients, whose subsequent study will occupy us in section~\ref{S:Long}. In section \ref{S:Short} we study the near field of the source using a different parametrization from the one given in \ref{SS:BoundaryStress}.

\subsection{Constraints on the source terms}
\label{SS:Source}

Consider a probe-source moving in the $AdS_5$-Schwarzschild background  \eqref{E:AdSBH}.  We require that the stress tensor of the probe-source
 $J_{\mu\nu}$ is conserved:
\begin{equation}
\label{E:DJeq0}
	D_{\mu}J^{\mu\nu} = 0 \,,
\end{equation}
where $D_\mu$ is the usual covariant derivative in curved spacetime.  If \eno{E:DJeq0} fails, the Einstein equations will be inconsistent. One way to see this inconsistency is to pass to an axial gauge where five of the fifteen independent components of the metric are set to zero.  There are nevertheless fifteen Einstein equations for the remaining ten components of the metric, and five of them are first order constraints whose consistency with the ten second order equations of motion depends on \eno{E:DJeq0} being true.  In terms of the Fourier coefficients $J_{\mu\nu}(\omega,\vec{k},z)$ introduced in \eno{FourierTransform}, the conservation equation \eqref{E:DJeq0} takes the form
\begin{subequations}
\label{E:DJeq0Explicit}
\begin{align}
\label{E:DJeq0Jm5}
	\frac{i\omega}{h}J_{m0}+i k_i J_{mi}&= - \frac{\left(J_{m5}\alpha^3 h\right)^{\prime}}{\alpha^3}\\
\label{E:DJeq0J55}
	\frac{i\omega}{h}J_{05}+i k_i J_{i5}&= J_{ii}\frac{\alpha^{\prime}}{\alpha}-J_{00}\frac{(h\alpha^2)^{\prime}}{2 h^2 \alpha^2}-\frac{(J_{55}\alpha^2 h^{3/2})^{\prime}}{\alpha^2 h^{1/2}} \,,
\end{align}
\end{subequations}
where we have defined
\begin{equation}
\label{E:Defofk}
	k_m = \begin{pmatrix} -\omega & k_1 & k_2 & k_3 \end{pmatrix}
\end{equation}
and used the labels $m=0,1,2,3$ and $i=1,2,3$.

Consider a near boundary series expansion of the Fourier coefficients $J_{\mu\nu}(\omega,\vec{k},z)$ in $z$,\footnote{In section \ref{S:Summary} we claimed that the existence of a series expansion of the Fourier coefficients in small $z$ is not required to obtain \eqref{E:Tmnsourced}. This is correct as can be seen by exchanging $i k_{m}$ with $\partial_{m}$ in the analysis leading to \eqref{E:NonCT} and \eqref{E:TraceT}. Nevertheless, we have chosen to work in Fourier space so that the results we obtain here may be easily referred to from later sections where we do require a series expansion in small $z$ for the Fourier coefficients of $J_{\mu\nu}$.}  
similar to \eno{Jseries},
\begin{equation}
 \label{Jseries2}
 J_{\mu\nu}(\omega,\vec{k},z) =
    \sum_{a=a_0}^\infty J^{(a)}_{\mu\nu}(\omega,\vec{k}) z^a \,.
\end{equation}
We will assume that $a$ runs over integers. Near the boundary $G_{\mu\nu} \sim 1/z^2$, so we must take $a_0 > -2$ in order for the contribution of $J_{\mu\nu}(\omega,\vec{k},z)$ to the Einstein equations to be less divergent near $z=0$ than the cosmological constant term ${12 \over L^2} G_{\mu\nu}$. If $a_0 \leq -2$, then generically the spacetime will no longer be asymptotically anti-de Sitter. Plugging \eno{Jseries} into the conservation equations \eno{E:DJeq0Explicit}, one finds that $J_{m5}^{(-1)}=0$, and that the first few coefficients $J_{\mu\nu}^{(a)}$ are related by the equations
\begin{align}
	ik^m  J_{nm}^{(a-1)}&=(3-a)J_{n5}^{(a)} \nonumber \\
	J_{55}^{(a-1)}(3-a)&=J^{m\,(a-1)}_m+i k^m J_{m5}^{(a-2)}
	\label{E:DJeq0Series}
\end{align}
where $a=0,\ldots,3$.  The components $J_{55}^{(2)}$ and $J_{n5}^{(3)}$ remain undetermined by these relations. They are integration constants of \eqref{E:DJeq0}. In \eqref{E:DJeq0Series} and in the rest of this work, the index $m$ is raised using the Minkowski metric.

\subsection{Perturbations of the five-dimensional metric}
\label{SS:MetricPerturbations}

In order to compute $\langle T_{mn} \rangle$ using the gauge-string duality we need to find the small $z$ asymptotics of the linear response of the metric to the source.  We write the perturbations as follows:
\begin{equation}
\label{E:Gmn}
    G_{\mu\nu} = G_{\mu\nu}^{\rm AdSBH}+\frac{\kappa_5^2 \alpha^2}{2 L^3} H_{\mu\nu} \,,
\end{equation}
where $G_{\mu\nu}^{\rm AdSBH} dx^{\mu}dx^{\nu}$ is the AdS${}_5$-Schwarzschild line element \eno{E:AdSBH}, and $H_{\mu\nu}$ is small compared to $G_{\mu\nu}^{\rm AdSBH}$.  The normalization factors in the second term in \eno{E:Gmn} are chosen so that if $H_{\mu\nu} = H_{\mu\nu}^{(4)} z^4 + {\cal O}(z^5)$ and we work in a gauge where $H_{\mu5}=0$, then the holographic stress tensor is
 \eqn{HolographicStress}{
  \langle T_{mn} \rangle = \langle T_{mn} \rangle_{\rm bath} + H_{mn}^{(4)} \,.
 }

To determine $H_{\mu\nu}$, one must solve the linearized version of the Einstein equations \eno{EinsteinEqs}, which we will write formally as
\begin{equation}
\label{E:DHeqJ}
	\mathcal{D}_{\mu\nu\rho\sigma}H^{\rho\sigma} = J_{\mu\nu} \,.
\end{equation}
The equations where $\mu$ and $\nu$ run from $0$ to $3$ comprise ten second order equations of motion,
and the equations with $\mu=5$ give five first order constraint equations.
The most general solution to the second order equations of motion involves twenty integration constants.  Ten set the value of $H_{mn}(0)$ and are related to deformations of the boundary theory metric. We insist that this metric should not be deformed, and so $H_{mn}(0)=0$. If we use the following expansion (c.f.~\cite{deHaro:2000xn}):
\begin{equation}
\label{E:HSeries}
	H_{mn}(z) = H_{mn}^{(1)} z + H_{mn}^{(2)} z^2 +
	  H_{mn}^{(3)} z^3 +
	  \tilde{H}_{mn}^{(4)} z^4 \ln z/L +
	  H_{mn}^{(4)} z^4 + \ldots \,,
\end{equation}
then the other ten integration constants are given by $H_{mn}^{(4)}$. The $H_{mn}^{(a)}$ with $a<4$, and also $\tilde{H}_{mn}^{(4)}$, may be determined by solving the second order equations of motion perturbatively in $z$. We find for $a=1,\,2$ or $3$,
\begin{subequations}
\label{E:HmnToJ}
\begin{equation}
\label{E:HmnToJ1}
	H_{mn}^{(a)}=
	 \frac{4}{a(4-a)}\left(J_{mn}^{(a-2)}-\frac{1}{3}\eta_{mn}J^{(a-2)\,s}_s
	-\frac{L^3}{\kappa_5^2}\left(R_{nm}^{(a-2)}-\frac{1}{6}R^{(a-2)} \eta_{nm}\right)
	\right) \,,
\end{equation}
and also
\begin{equation}
	\tilde{H}_{mn}^{(4)}=
	-\left(J_{mn}^{(2)}-\frac{1}{3}\eta_{mn}J^{(2)\,s}_s
	-\frac{L^3}{\kappa_5^2}\left(R_{nm}^{(2)}-\frac{1}{6}R^{(2)} \eta_{nm}\right)\right) \,,
\end{equation}
\end{subequations}
where $R_{nm}$ and $R$ are the Ricci Scalar and Ricci tensor which follow from the metric $g_{mn} = \eta_{mn}+\frac{\kappa_5^2}{2 L^3}H_{mn}$ expanded to linear order in $H_{mn}$. Following the notation in \eqref{Jseries} and \eqref{E:HSeries}, we have defined
\begin{align}
	R_{mn}&=R_{mn}^{(1)}z+R_{mn}^{(2)}z^2+\ldots\\
	R&=R^{(1)}z+R^{(2)}z^2+\ldots.
\end{align}
So, for example, $R_{mn}^{(1)}$ is a linear combination of second derivatives of $H_{mn}^{(1)}$.

Once the $H_{mn}$ are known, we can compute the boundary energy momentum tensor by varying the action with respect to the boundary value of the metric. We carry out this procedure in appendix \ref{A:HoloRG}. The resulting stress tensor takes the form
\begin{equation}
\label{E:Tmntotal}
	\langle T_{mn} \rangle
	=
	\langle T_{mn}\rangle_{\rm bath}
	+\langle T_{mn}^{\epsilon} \rangle
	+\langle \delta T_{mn} \rangle\,,
\end{equation}
where 
$\langle \delta T_{mn} \rangle$ is the finite part of the fluctuation of the stress-tensor above the background value of the plasma and
$\langle T_{mn}^{\epsilon} \rangle$ is a divergent contribution to the stress energy tensor, whose explicit form is given in \eqref{E:Tmnepsilon}.
Usually, divergent terms in the stress energy tensor can be removed by introducing appropriate counter-terms via holographic renormalization. The infinities which appear in \eqref{E:Tmntotal} are of a different nature: we expect that they are associated with the probe source having parameters which formally diverge. 
For example, string configurations which end on the asymptotically AdS boundary are dual to infinitely massive quarks. We show in section \ref{S:Examples} that all the divergent terms in the boundary theory stress tensor $\langle T_{mn}^{\epsilon} \rangle$ are a result of the infinite mass of the quark.
More generally, to the extent that the $J_{mn}$ are localized in space, the divergent terms in $\langle T_{mn}^\epsilon \rangle$ are likewise localized, meaning that we may treat them as divergent contact terms supported at the location of the probe.  See appendix \ref{A:HoloRG} for an extended discussion of these divergences.

The non-divergent part of the fluctuation of the stress-energy tensor above the background value of the plasma is given by
\begin{equation}
\label{deltaTmn}
	\langle \delta T_{mn} \rangle = H_{mn}^{(4)} - \eta_{mn}H^{(4)\,s}_s + \frac{3 L^3}{4\kappa_5^2}\left(R_{mn}^{(2)}-\frac{1}{2}R^{(2)}\eta_{mn}\right)-\frac{1}{4}J_{mn}^{(2)} \,.
\end{equation}

\subsection{Constraints on the boundary stress tensor}
\label{SS:BoundaryStress}

To evaluate $\langle \delta T_{mn} \rangle$, we need to calculate the $H_{mn}^{(4)}$'s in \eno{E:HSeries}. Five of the ten coefficients $H_{mn}^{(4)}$ can be determined by perturbatively solving the five first order constraint equations. From the $55$ equation,
 \eqn{E:Trace}{
  H^{m\,(4)}_m = {1 \over 6} J^{m\,(2)}_m -
    {1 \over 3} J_{55}^{(2)} \,,
 }
and from the $m5$ equations,
 \eqn{E:Constraint}{
  i k^m H_{mn}^{(4)} = J_{n5}^{(3)} + i k_n H^{m\,(4)}_m \,,
 }
where we made use of \eqref{E:DJeq0Series} and \eqref{E:HmnToJ}. Had the conservation condition \eqref{E:DJeq0} not been satisfied, the constraint equations would have been inconsistent with the second order equations of motion.
In light of the constraints \eno{E:Trace} and \eqref{E:Constraint}, there are only five independent components of $H_{mn}^{(4)}$.  They are determined by solving the Einstein equations with infalling boundary conditions at the horizon. We will explain how to do this explicitly in section~\ref{S:Long} in the context of a small $k$ expansion (and in section~\ref{S:Short} for a large $k$ expansion.)  Here we wish to avoid any use of an expansion in small or large $k$, but we will solve the equations \eno{E:Trace} and~\eno{E:Constraint} in a way that will make it as easy as possible to study a small $k$ expansion later.  A good strategy is to choose the undetermined components to be the traceless spatial part of $H_{mn}$, which we denote as $H_{(ij)}$.  The reason this is a good strategy is that when hydrodynamics is valid, the components $\langle \delta T_{(ij)} \rangle$ of the stress tensor are usually one order in $k$ smaller than the other components for small $k$: they arise from viscous effects in the regime of linearized hydrodynamics.  It is straightforward to use \eno{E:Trace},~\eno{E:Constraint}, and
\begin{equation}
\label{E:R2properties}
	\frac{L^3}{\kappa_5^2}R^{(2)} = -J^{(2)m}_{m} \quad\qquad i k^m \frac{L^3}{\kappa_5^2} R_{mn}^{(2)} = -\frac{1}{2}i k_n J_{m}^{(2)\,m} \,,
\end{equation}
together with~\eno{deltaTmn}, to show that
\begin{align}
\label{E:NonCT}
	i k^m \langle \delta T_{mn} \rangle &= J_{n5}^{(3)}\\
\label{E:TraceT}
	\langle \delta T_{m}^m \rangle & = J_{55}^{(2)}.
\end{align}
For completeness, we note that the divergent terms satisfy
\begin{align}
	i k^m \langle T_{mn}^{\epsilon} \rangle &= \frac{J_{n5}^{(0)}}{\epsilon^3}+\frac{J_{n5}^{(1)}}{\epsilon^2}+\frac{J_{n5}^{(2)}}{\epsilon}\\
	\langle T_{m}^{\epsilon\,m} \rangle & =
	 \frac{J_{55}^{(-1)}}{\epsilon^3}+\frac{J_{55}^{(0)}}{\epsilon^2}+\frac{J_{55}^{(1)}}{\epsilon} \,,
\end{align}
where we made use of
\begin{equation}
\label{E:R1properties}
	\frac{L^3}{\kappa_5^2}R^{(1)} = 2(J_{55}^{(1)}-J^{(1)m}_{m})\quad\qquad i k^m \frac{L^3}{\kappa_5^2} R_{mn}^{(1)} = i k_n (J_{55}^{(1)}-J^{(1)m}_{m})
\end{equation}
and \eqref{E:DJeq0Series}.

Using \eqref{E:NonCT} and \eqref{E:TraceT} we find
\begin{multline}
\label{E:Tmn}
    \langle \delta T_{mn} \rangle =
    \frac{-i\omega J_{05}^{(3)} + i k_i J_{i5}^{(3)}+k_i k_j
      H_{(ij)}^{(4)}}{\vec{k}^2-3\omega^2}
    \begin{pmatrix}
	-3
		& \frac{k_1}{\omega}
		& \frac{k_2}{\omega}
		& \frac{k_3}{\omega} \\
	\frac{k_1}{\omega}
		& -1
		& 0 & 0 \\
	\frac{k_2}{\omega}
		& 0
		& -1
		& 0 \\	
	\frac{k_3}{\omega}
		& 0
		& 0
		& -1 \\
	\end{pmatrix}
    \\
    -i\frac{J_{i5}^{(3)}-i k_j
     H_{(ij)}^{(4)} }{\omega}
    \begin{pmatrix}
        0 & \delta_{i1} & \delta_{i2} & \delta_{i3} \\
        \delta_{i1} & 0 & 0 & 0\\
        \delta_{i2} & 0 & 0 & 0\\
        \delta_{i3} & 0 & 0 & 0\\
    \end{pmatrix}
    +
	\begin{pmatrix}
	0 & \phantom{0} & 0 & \phantom{0} \\
	\phantom{0} & \phantom{0} & \phantom{0} & \phantom{0} \\
	0 & \phantom{0} & H_{(ij)}^{(4)} & \phantom{0} \\
	\phantom{0} & \phantom{0} & \phantom{0} & \phantom{0} \\
	\end{pmatrix}
    +
	\mathcal{A}_{mn}
\end{multline}
where the tensor ${\cal A}_{mn}$ is given by
\begin{multline}
 \label{E:TraceAnomaly}
	\mathcal{A}_{mn}=
    \frac{k_i k_j
      \mathcal{R}_{ij}}{\vec{k}^2-3\omega^2}
    \begin{pmatrix}
	-3
		& \frac{k_1}{\omega}
		& \frac{k_2}{\omega}
		& \frac{k_3}{\omega} \\
	\frac{k_1}{\omega}
		& -1
		& 0 & 0 \\
	\frac{k_2}{\omega}
		& 0
		& -1
		& 0 \\	
	\frac{k_3}{\omega}
		& 0
		& 0
		& -1 \\
	\end{pmatrix}
    -\frac{k_j \mathcal{R}_{ij} }{\omega}
    \begin{pmatrix}
        0 & \delta_{i1} & \delta_{i2} & \delta_{i3} \\
        \delta_{i1} & 0 & 0 & 0\\
        \delta_{i2} & 0 & 0 & 0\\
        \delta_{i3} & 0 & 0 & 0\\
    \end{pmatrix}+
    	\begin{pmatrix}
	0 & \phantom{0} & 0 & \phantom{0} \\
	\phantom{0} & \phantom{0} & \phantom{0} & \phantom{0} \\
	0 & \phantom{0} & \mathcal{R}_{ij} \\
	\phantom{0} & \phantom{0} & \phantom{0} & \phantom{0} \\
	\end{pmatrix}
\\
	+\frac{\langle \delta T^l_l \rangle}{\vec{k}^2-3\omega^2}
	\begin{pmatrix}
	-\vec{k}^2
		& k_1 \omega
		& k_2 \omega
		& k_3 \omega\\
	k_1 \omega
		& -\omega^2
		& 0 & 0 \\
	k_2 \omega
		& 0
		& -\omega^2
		& 0 \\
	k_3 \omega
		& 0 & 0
		& -\omega^2
	\end{pmatrix}
    \,.
\end{multline}
where
\begin{equation}
	\mathcal{R}_{ij} = \frac{1}{4}\left(\frac{3 L^3}{ \kappa_5^2} R_{(ij)}^{(2)}-J_{(ij)}^{(2)}\right)
\end{equation}
and the sum over $l$ in $\langle \delta T^l_l \rangle$ runs from $0$ to $3$. We have included the $\mathcal{R}_{ij}$ terms together with the trace anomaly term (the last term on the right hand side of \eqref{E:TraceAnomaly}) since the former will also induce a $\ln \epsilon/L$ divergence in the stress tensor which is indicative of a conformal anomaly \cite{deHaro:2000xn}.
As observed before, we did not use a small $k$ expansion to derive \eno{E:Tmn}: in this sense it is non-perturbative.

Suppose one is working in a regime where linearized hydrodynamics applies (i.e.~small enough $k$).
The first term in \eno{E:Tmn} then relates to sound modes.  The pole structure $1/(\vec{k}^2-3\omega^2)$ gives rise to a Mach cone in real space if the source is moving supersonically.  The second term in \eno{E:Tmn}, with the $1/\omega$ pole structure, is associated with diffusion of energy in the plasma, and in a suitable setup it signals the formation of a laminar wake far behind the source. It is commonplace for these poles to be shifted to slightly imaginary values by viscous effects, see equations \eqref{E:resummed}. Either the $\omega=\pm\sqrt{3} |\vec{k}|$ poles or the $\omega=0$ pole could be canceled by an appropriate numerator at leading order.  If this happens, it corresponds to suppression of the sonic boom or the diffusion wake.

In the near field, where the hydrodynamic approximation is not valid, we 
find that the $H_{(ij)}^{(4)}$ completely cancel the pole structure in \eqref{E:Tmn} and therefore $H_{(ij)}$ are not such good variables.
In fact, in \cite{Kovtun:2005ev,Gubser:2007nd,Chesler:2007sv} it was shown how one may choose combinations of the components of the metric fluctuations whose equations of motion are completely decoupled from each other. While this is convenient when studying the solutions to the equations of motion at all scales, we find that the current choice shows the relation to hydrodynamics in a more transparent way.

\section{Large distance asymptotics of $\langle \delta T_{mn} \rangle$}
\label{S:Long}

As we have already remarked, if linearized hydrodynamics applies, $\langle \delta T_{(ij)} \rangle$ is expected to be subleading compared to other components of $\langle \delta T_{mn} \rangle$ by a single factor of $k$, because $\langle \delta T_{(ij)} \rangle$ has to do with viscous effects.
If we set $H_{(ij)}^{(4)}=0$ in \eqref{E:Tmn} then at leading order in $k$ the stress tensor is completely determined
by the non-conservation equation \eno{E:Tmnsourced} together with tracelessness (assuming  no anomaly term $\mathcal{A}_{mn}$). As a result, the constitutive relations of inviscid linearized hydrodynamics hold, likewise at leading order in $k$.
One might think that the next correction in small $k$ comes entirely from improving the constitutive relations by adding shear viscosity.  What this section aims to show is that this is not quite right: in general, there is a non-hydrodynamical correction term ${\cal F}_{mn}$,
as described in \eno{MainResult}. It enters at the same order as viscous effects if $J_{(ij)}$ is of the same order in $k$ as the four-force density $J_{m5}^{(3)}$.

To find $H_{(ij)}^{(4)}$, we need to consider some of the second order equations of motion, which appear in appendix~\ref{A:EOM}.
\begin{subequations}\label{E:REagain}
 \begin{align}
	\partial_Y^2 H_{(ij)} + \left(\frac{z_0 \omega}{4}\right)^2 \left(1-e^Y\right)^{-3/2} H_{(ij)} &= -\frac{z_0^2}{4} J_{(ij)} \left(1-e^Y\right)^{-3/2} e^Y  \nonumber \\ &\qquad{} -\frac{z_0^2 \omega}{8} k_{(i}H_{j)0}\left(1-e^Y\right)^{-3/2} + \mathcal{O}(k_i^2 H_{mn})  \label{E:HijEOM} \\
  	\partial_Y^2 H_{0i} - \partial_Y H_{0i} &= -\frac{z_0^2}{4}J_{0i}(1-e^Y)^{-3/2}e^Y+\mathcal{O}(k_i^2 H_{mn})  \nonumber \\
	&\qquad{} + \mathcal{O}(\omega k_i H_{mn}) \,,\label{E:H0iEOM}
 \end{align}
 \end{subequations}
where we have defined a new variable
\begin{equation}
\label{YDef}
    Y = \ln h \,.
\end{equation}
In the $Y$ variable, the boundary is located at $Y=0$ and the horizon at $Y=-\infty$.
The notation ${\cal O}(k_i^2 H_{mn})$ in \eno{E:REagain} means any combination of two factors of spatial components $k_i$ with one factor of a component of the metric perturbation: for example, $k_1 k_2 H_{00}$ would be such a term.  Similarly, ${\cal O}(\omega k_i H_{mn})$ in \eno{E:H0iEOM} means any combination of a factor of $\omega$, a component of the spatial momentum $k_i$, and a component of the metric perturbation.  In contrast, by ${\cal O}(k)$ we mean a term containing one factor of $\omega$ or $k_i$.

We claim that in order to find $\langle \delta T_{mn} \rangle$ to an accuracy of one factor of $k$ beyond the inviscid linearized hydro approximation, precisely the terms shown in \eno{E:REagain} need to be retained.  This may seem counter-intuitive given that we have dropped factors quadratic in $k_i$ from \eno{E:REagain} while keeping factors quadratic in $\omega$.  We will see in section section~\ref{SS:Drag} that this seemingly uneven retention of ${\cal O}(k^2 H_{mn})$ terms is exactly what is required, because of the structure of the event horizon.

\subsection{Inviscid hydrodynamics and the non-hydrodynamical correction}
\label{SS:Naive}

Equations \eqref{E:Tmn} and \eqref{E:REagain}, together with the assumption that the plasma is at rest at infinity, teach us that $H_{mn}$ must be $\mathcal{O}(k^{-n}J_{\mu\nu})$ with $n \geq -1$ (we must allow $n=-1$ due to the constraint equations \eqref{E:Constraint}).  Thus, to order $\mathcal{O}(k^{0}J_{\mu\nu})$, one may neglect ${\cal O}(\omega^2 H_{ij})$ and ${\cal O}(\omega k_i H_{0j})$ terms, and \eno{E:HijEOM} becomes
 \eqn{LeadingOrder_zvars}{
  \partial_Y^2 H_{(ij)} = -{z_0^2 \over 4} J_{(ij)} (1-e^Y)^{-3/2}
    e^Y \,,
 }
whose solution is
 \eqn{IntegrateTwice}{
   H_{(ij)} = Q_{ij} Y + {z_0^2 \over 4}
    \int_Y^0 dy \int_{-\infty}^y dy' \, J_{(ij)}(y') (1-e^{y'})^{-3/2}
      e^{y'} \,.
 }
Now let's ask what happens when we plug the small $z$ series expansions \eno{Jseries} and~\eno{E:HSeries} into \eno{IntegrateTwice} and compare term by term.  By comparing up to ${\cal O}(z^4 \log z)$ in \eno{IntegrateTwice}, one can recover the terms in \eno{E:HmnToJ} that do not involve the Ricci scalar or tensor (more precisely, one can recover the $mn = (ij)$ components of those terms.)  These divergent terms are dealt with in section~\ref{S:EMTensor} and appendix~\ref{A:HoloRG}.  The ${\cal O}(z^4)$ term in \eno{IntegrateTwice} gives us
 \eqn{GotHijFour}{
  H^{(4)}_{(ij)} =
   -{Q_{ij} \over z_0^4} +{\cal F}_{ij} \,,
 }
where ${\cal F}_{ij}$ is defined in \eno{FResult}.

To determine the integration constant $Q_{ij}$, we need to impose boundary conditions at the horizon which permit infalling waves but not outgoing ones.  In the notation we're using, this implies that we need to retain terms which oscillate like $e^{-\frac{z_0}{4}i\omega Y} \sim 1 -\frac{z_0}{4}i\omega Y$ but not those which oscillate like $e^{+\frac{z_0}{4}i\omega Y}$. This will be carried out and explained in more detail in section \ref{SS:Drag}.  As a lowest order approximation (valid formally to order $\mathcal{O}(k^{-1}J_{m5})$) permitting only infalling modes implies that $H_{(ij)}$ asymptotes to a constant.
Thus, according to \eqref{IntegrateTwice}, at this order $Q_{ij}=0$ which, together with \eqref{E:Tmn}, gives us inviscid hydro up to the $\mathcal{F}_{ij}$ terms and the trace anomaly term $\mathcal{A}_{mn}$.  Corrections to $Q_{ij}$ at higher orders in small $k$ will lead to viscous hydrodynamics. These will be dealt with in section \ref{SS:Drag}.

\subsection{Small $k$ properties of $J_{\mu\nu}(\omega,\vec{k},z)$}
\label{SS:SmallK}

Before proceeding with an analysis of the near horizon boundary conditions, we make an aside to discuss in more detail what we mean by a small $k$ expansion.
Let's introduce the following notation for expansions in small $k$ and small $\omega$:
\begin{equation}
	\label{E:JkSeries}
	J_{\mu\nu}(\omega,\vec{k}) = \sum_{\overline{a}=\overline{a_0}}^{\infty}J_{\mu\nu}^{(\overline{a})}(\omega,\vec{k})
\end{equation}
where
\begin{equation}
\label{E:JkSeriesB}
	J_{\mu\nu}^{(\overline{a})}(\lambda\omega,\lambda\vec{k})
	= \lambda^{\overline{a}} J_{\mu\nu}^{(\overline{a})}(\omega,\vec{k})\,.
\end{equation}
In \eqref{E:JkSeries} and~\eqref{E:JkSeriesB} we have suppressed the dependence on the radial coordinate $z$. In some instances $J_{\mu\nu}$ will have delta-function support in Fourier space, meaning that $\omega$ will be localized around some $\omega_0(k)$.  This may happen when $J_{\mu\nu}(t)$ is not localized in time.  To keep the discussion general, we consider a source whose stress energy tensor can be factored into an analytic and a non-analytic term:
\begin{equation}
\label{E:JtoJDelta}
    J_{\mu\nu}(\omega,\vec{k},z) = \Delta(\omega,\vec{k})
      j_{\mu\nu}(\omega,\vec{k},z) \,,
\end{equation}
where $\Delta(\omega,\vec{k})$ contains any non-analyticity. If $J_{\mu\nu}(\omega,\vec{k},z)$ are analytic in $\omega$ and $\vec{k}$, we may simply set $\Delta = 1$.
The assumption that the probe source is localized in space amounts to having a smooth $j_{\mu\nu}(\omega,\vec{k},z)$ on the right hand side of \eqref{E:JtoJDelta}: that is, $j_{\mu\nu}(\omega,\vec{k},z)$ can be expanded in a Taylor series around $k=0$.
Actually, the definition of $j_{\mu\nu}$ as presented in \eqref{E:JtoJDelta} is somewhat ambiguous since we may always rescale $\Delta$ and $j_{\mu\nu}$ by appropriate factors of the momentum, for example, we may take $\Delta \to \Delta/\omega$ and $j_{\mu\nu} \to \omega j_{\mu\nu}$. To fix this ambiguity we require that $j_{\mu\nu}$ is of order $k^n$ with the smallest possible $n$.
As noted in section \ref{S:Summary} one may easily generalize these results to the case where $J_{\mu\nu}$ is a sum of terms, each of the form shown on the right hand side of \eno{E:JtoJDelta}.

The superscripts $n$ and $\overline{n}$ will be used fairly often in the following sections, not only for $j_{\mu\nu}$ but for other functions as well. Thus, $f^{(n,\overline{m})}$ specifies the $n$'th component of $f(\omega,\vec{k},z)$ in a series expansion near the AdS boundary, and the $m$'th component of $f^{(n)}(\omega,\vec{k},z)$ in a small momentum/frequency expansion.

Shifting our attention to the near horizon asymptotics of $J_{mn}$,
we require that the stress-energy tensor of the source is causal in the sense that the (linear) response of the metric to the source will not be forced to have modes outgoing from the horizon, and that it is non-divergent.

\subsection{One order beyond inviscid hydrodynamics}
\label{SS:Drag}

If the leading behavior of $J^{(3)}_{m5}(\omega,\vec{k})$ is of the same order or lower order (in a small momentum expansion) than the leading behavior of ${\mathcal{F}}_{(ij)}$, then we can go one order beyond inviscid hydrodynamics and obtain the first viscous correction to it. For simplicity we assume that the leading behavior of $j^{(3)}_{m5}$ is of order $k^0$, as is the leading behavior of $\mathcal{F}_{ij}/\Delta$. Our results may be easily generalized to scenarios where $j^{(3)}_{m5}$ starts at an order $k^n$ greater than that of $\mathcal{F}_{ij}/\Delta$. We will comment on this generalization where appropriate.
Our strategy to obtain the inviscid corrections
is to compare a small momentum/frequency expansion of $h_{mn}(z)$, where $H_{mn}(z) = \Delta h_{mn}(z)$,
to a near horizon expansion of the solution.

A small momentum/frequency expansion of $h_{(ij)}$ can be read off from \eqref{IntegrateTwice}:
\begin{equation}
\label{E:Smallk}
    h_{(ij)} = q_{ij}Y+\frac{z_0^2}{4}\int_Y^0 dy \int_{-\infty}^y dy^{\prime}j_{(ij)}(y')(1-e^{y^{\prime}})^{-3/2}e^{y^{\prime}} \,,
\end{equation}
where we have used
$
    Q_{ij} = q_{ij} \Delta
$.
To find $q_{ij}$, we consider the near horizon asymptotics of \eqref{E:REagain} (with $H_{mn}$ replaced by $h_{mn}$ and $J_{mn}$ replaced with $j_{mn}$).
It may be intuitively helpful to think of \eno{E:REagain} as describing coupled oscillators evolving in a ``time'' parametrized by the coordinate $Y$ (although in the $AdS_5$-Schwarzschild geometry, $Y$ is a spacelike coordinate) and perturbed by ``forces'' corresponding to the $J_{mn}$ terms.  Near the horizon, the coefficient of $h_{(ij)}$ in \eno{E:HijEOM} is constant, so $h_{(ij)}$ ``oscillates'' with frequency $z_0\omega/4$.  On other other hand, the $h_{0i}$ equations have no ``restoring force'' term, and the sign of the $\partial_Y h_{0i}$ term is such that the ``motion'' of $h_{0i}$ is damped as one approaches the horizon.  Thus the $h_{0i}$ tend to a finite limit as $Y\to -\infty$:
 \eqn{PoyntingLimit}{
  \lim_{Y \to -\infty} h_{0i}(\omega,\vec{k},Y) =
    -W_j(\omega,\vec{k})
 }
for some $W_j(\omega,\vec{k})$.
The ${\cal O}(k_i^2 h_{mn})$ terms that were dropped from \eno{E:H0iEOM} do not affect the conclusion \eno{PoyntingLimit} because they are suppressed by a factor of $e^Y$, and because all the $h_{mn}$ must be assumed to be bounded as $Y \to -\infty$ in order for the linearized analysis to be valid.

Using \eqref{PoyntingLimit} the near horizon behavior of $h_{(ij)}$ takes the form
\begin{equation}
\label{E:Nearh}
	h_{(ij)} = U_{ij} e^{-\frac{1}{4}iz_0 \omega Y}+V_{ij} e^{\frac{1}{4}iz_0 \omega Y}
        +\frac{2k_{(i}W_{j)}}{\omega} + \left(\substack{\hbox{terms which vanish} \\ \hbox{at the horizon}}\right)
        +\mathcal{O}(k) \,,
\end{equation}
where $U_{ij}$ and $V_{ij}$ are integration constants and we have assumed that $J_{(ij)} e^Y$, and so also $j_{(ij)} e^Y$, vanish at the horizon.  Standard horizon boundary conditions are to allow only infalling modes: thus $V_{ij} = 0$.

Comparing the small momentum limit of the near horizon asymptotics \eqref{E:Nearh} with the near horizon asymptotics of the small momentum limit \eqref{E:Smallk} implies that
\eqn{MatchingComparison}{
  h_{(ij)} = U_{ij} \left( 1 - {1 \over 4} i z_0 \omega Y \right) +
    {2k_{(i} W_{j)} \over \omega} + \mathcal{O}(k) =
   q_{ij} Y + \left(\substack{\hbox{Finite corrections}\\ \hbox{of order }k^0}\right) +\mathcal{O}(k)\,.
 }
In \eno{MatchingComparison} we made use of the method of matched asymptotic expansions: we start with \eno{E:Smallk}, which is valid for momenta small enough that the oscillating term at the horizon can be Taylor expanded, $-{1 \over z_0 \omega} \ll Y$, and \eno{E:Nearh} which is valid near the horizon, $Y \ll -1$.  The matching region is the intersection of these two intervals:
 \eqn{MatchingRegion}{
  -{1 \over z_0 \omega} \ll Y \ll -1 \,.
 }
Provided $z_0 \omega \ll 1$, which is to say $\omega \ll T$, the region \eno{MatchingRegion} is non-empty.  
By matching the constant ($Y$-independent) terms in \eno{MatchingComparison} we find
 \eqn{MatchConstants}{
  U_{ij}^{(\overline{-1})} = -{2 k_{(i} W_{j)}^{(\overline{-1})} \over \omega}\,.
}
By matching the terms proportional to $Y$ in \eno{MatchingComparison} we find
\begin{align}
\label{MatchDerivatives}
  q_{ij}^{(\overline{-1})}&=0\\
  q_{ij}^{(\overline{0})} &= -{1 \over 4} i z_0 \omega U_{ij}^{(\overline{-1})} =
    {1 \over 2} i z_0 k_{(i} {W^{(\overline{-1})}_{j)}} \,,
\end{align}
where in the second step we used \eno{MatchConstants}.

To determine $W_{j}^{(\overline{-1})}$, we consider the small momentum/frequency expansion of the equation of motion for $h_{0i}$, \eqref{E:H0iEOM} with $H_{mn}$ replaced with $h_{mn}$ and $J_{mn}$ replaced with $j_{mn}$.
From \eqref{E:Tmn}, the leading small momentum contribution to $s_i$, defined through the Poynting vector $s_i \Delta  = S_i = -\langle T_{0i} \rangle $, is of order $\mathcal{O}(k^{-1})$. Thus,
 \eqn{E:SolH0i}{
  h_{0i} = s_i^{(\overline{-1})} (e^Y-1) +
    {\cal O}(k^0) \,,
 }
which gives us
 \eqn{WversusS}{
  W_j = s_j^{(\overline{-1})} + {\cal O}(k^0) \,.
 }
The traceless space-space components of the stress tensor in the absence of a trace anomaly, $H_{(ij)}^{(4)}$,
can be read off of \eqref{WversusS}, \eqref{MatchDerivatives} and \eqref{E:Smallk}:
\begin{equation}
\label{E:GotTij}
    H_{(ij)}^{(4)} = -\frac{1}{2\pi T}i k_{(i}S_{j)} +\mathcal{F}_{ij}
	+\left(\hbox{corrections}\right) \,,
\end{equation}
where we have used $z_0 = 1/\pi T$.

Plugging \eqref{E:GotTij} into \eqref{E:Tmn} we obtain \eqref{MainResult}, up to the aforementioned anomaly terms. The explicit values for the energy density $\epsilon = \Delta \left(\epsilon^{(\overline{-1})}+\epsilon^{(\overline{0})}\right)$ and Poynting vector $S = \Delta \left(s^{(\overline{-1})}+s^{(\overline{0})}\right)$ are
\begin{subequations}
\label{E:HydroEandS}
\begin{align}
\label{E:epsilonDm1}
 	\epsilon^{(\overline{-1})}\Delta &= -3\frac{i k_i j_{i5}^{(3,\overline{0})}-i\omega j_{05}^{(3,\overline{0})}}{\vec{k}^2-3\omega^2}\Delta\\
\label{E:epsilonD0}
 	\epsilon^{(\overline{0})}\Delta &=
	-3\frac{i k_i j_{i5}^{(3,\overline{1})}\Delta-i\omega j_{05}^{(3,\overline{1})}\Delta+k_i k_j \mathcal{F}_{ij}-i k^2 k_j s_j^{(\overline{-1})}\Delta/3\pi T}{\vec{k}^2-3\omega^2}\\
\label{E:SDm1}	
	s_i^{(\overline{-1})}\Delta &= i\frac{j_{i5}^{(3,\overline{0})}\Delta}{\omega} + \frac{i k_i j_{05}^{(3,\overline{0})} - i k_j j_{j5}^{(3,\overline{0})} k_i/\omega}{\vec{k}^2-3\omega^2}\Delta
\end{align}\vskip-0.4in
\begin{multline}
\label{E:SD0}
 	s_i^{(\overline{0})}\Delta = i\frac{
		j_{i5}^{(3,\overline{1})}\Delta
		-i k_j \mathcal{F}_{ji}
		-k_i k_j s_j^{(\overline{-1})}\Delta/12\pi T
		-k^2 s_i^{(\overline{-1})}\Delta/4\pi T }
	{\omega} \\
	+ \frac{i k_i j_{05}^{(3,\overline{1})}\Delta
	- \left(i k_j j_{j5}^{(3,\overline{1})}\Delta + k_i k_j \mathcal{F}_{ij} - i k^2 k_j s_j^{(\overline{-1})}\Delta/3\pi T \right)k_i/\omega}{\vec{k}^2-3\omega^2}
\end{multline}
\end{subequations}
Several remarks are in order. First, if the leading order behavior of $j_{m5}^{(3)}$ is of order $k^n$ then the superscripts $(\overline{-1})$ and $(\overline{0})$ should be replaced with $(\overline{n-1})$ and $(\overline{n})$, where it is understood that $s^{(\overline{m})}=0$ and $\epsilon^{(\overline{m})}=0$ for $m\leq n-2$. Also, if the leading behavior of $\mathcal{F}_{ij}$ is of higher order in $k$ than that of $J_{m5}$ then its contribution to \eqref{E:HydroEandS} should be dropped.
Second, we note that to leading order, a wake will exist as long as $\Delta$ does not cancel the $1/\omega$ pole of $s_i^{(\overline{-1})}$.

The expressions in \eno{E:epsilonD0} and \eqref{E:SD0} include terms behaving as $1/\omega^2$ and $1/(\vec{k}^2-3\omega^2)^2$ (coming, for example, from substituting the explicit expression for $s_j^{(\overline{-1})}$ into \eno{E:epsilonD0}).  These terms can be understood as shifting the position of the sound and diffusion poles to slightly imaginary frequency, indicating viscous attenuation.  Retaining $\mathcal{O}(k)$ accuracy and using
\begin{equation}
\label{E:Fhydro}
	F^{\rm hydro}_m = J_{m5}^{(3)}-i k^n \mathcal{F}_{nm}\,,
\end{equation}
we find the following resummations of \eno{E:HydroEandS}:
\begin{subequations}
\label{E:resummed}
\begin{align}
	\epsilon &= \frac{i k_i F^{\rm hydro}_i - i \omega F^{\rm hydro}_0 +\frac{1}{3\pi T}k^2 F^{\rm hydro}_{0}}{\omega^2 - \frac{1}{3}k^2 + \frac{1}{3\pi T}ik^2\omega} + \mathcal{O}(k J_{\mu\nu})\\
	S &= \frac{-\frac{1}{3}i k_i F^{\rm hydro}_0+i \omega k_i k_j F^{\rm hydro}_j/k^2}{\omega^2 - \frac{1}{3}k^2 + \frac{1}{3\pi T}ik^2\omega}-\frac{F^{\rm hydro}_i - k_i k_j F^{\rm hydro}_j/k^2}{i\omega-\frac{1}{4\pi T}k^2}+\mathcal{O}(k J_{\mu\nu})\,,
\end{align}
\end{subequations}
which are
precisely the expressions for the energy density and Poynting vector for a conformal fluid with $\eta/s = 1/4\pi$ in the linear hydrodynamic approximation, sourced by an effective hydrodynamic four-force density $F^{\rm hydro}$. The effective four force density $F^{\rm hydro}$ deviates from the actual four force $J_{m5}^{(3)}$ by the $k^n \mathcal{F}_{nm}$ term on the right hand side of \eqref{E:Fhydro}. It would be interesting to see how higher order corrections arise \cite{Bhattacharyya:2007jc,Baier:2007ix}.  Obtaining \eno{E:resummed} completes our demonstration of \eno{MainResult} in the case where $J^{(3)}_{m5}$ is of the same or lower order in a small $k$ expansion than ${\cal F}_{ij}$.

From our construction, we can see how the constitutive relations arise from the Einstein equations. Near the horizon, the traceless space-space components of the metric are coupled to the $0i$ components of the metric in such a way that imposing no outgoing modes from the horizon induces the constitutive relations among the remaining integration constants.

\subsection{Configurations without a wake}
\label{SS:NoDrag}

In contrast to the discussion in section~\ref{SS:Drag}, let us consider the case where the leading behavior of $\mathcal{F}_{ij}$ at small $k$ dominates over the leading behavior of $J_{m5}^{(3)}$.  Then \eqref{GotHijFour} and the discussion following it implies that
\begin{equation}
\label{E:LeadingF}
	H_{(ij)}^{(4)} = \mathcal{F}_{ij} + \mathcal{O}(J_{m5}^{(3)})\,.
\end{equation}
The $\mathcal{O}(J_{m5}^{(3)})$ corrections now become more difficult to evaluate since the near horizon behavior of $H_{0i}$ is hard to access in this setting. At this point, our perturbative analysis allows us to study only the inviscid limit of such configurations.
If the leading behavior of $\mathcal{F}_{ij}/\Delta$ is of order $k^0$ then the leading terms in an asymptotic expansion of the energy density and Poynting vector around $k=0$ can be read off of \eqref{E:HydroEandS} after setting $j_{m5}^{(3,\overline{0})}$ to zero. Otherwise, if $\mathcal{F}_{ij}$ is $\mathcal{O}(k^n\Delta)$, we need to replace $(\overline{0})$ and $(\overline{1})$ with $(\overline{n})$ and $(\overline{n+1})$ and set $j_{m5}^{(3,\overline{n})}=0$. As in the previous section, we will assume for simplicity that $\mathcal{F}_{ij}$ is of order $k^0\Delta$. The generalization to a higher order dependence on the momentum is straightforward.

Let's look for the conditions for the absence of a wake, or a $1/\omega$ pole in the leading terms for the Poynting vector, once $J_{m5}^{(3)}$ is subleading.
If $\Delta$ doesn't introduce such a pole, then this could happen if $k^i\mathcal{F}_{ij}$ has an $\omega$ dependence which will cancel the $1/\omega$ poles in \eqref{E:HydroEandS}:
using \eqref{E:DJeq0Explicit} we can rewrite $k_i \mathcal{F}_{ij}$ in terms of integrals over the $J_{0i}$'s and over $J^{j}_{j}$,
\begin{align}
\label{E:dTi}
	i k_j \mathcal{F}_{ij} & = - {\rm F} \left[\int_{z}^{z_0} 
	 d\zeta \left( \partial_\zeta
        \left(J_{i5}h\zeta^{-3}\right)+ i \omega \frac{J_{0i}\zeta^{-3}}{h} +\frac{1}{3} i k_i J_{j}^{j} \zeta^{-3} \right) \right] \\
    &= -i \omega {\rm F} \left[\int_{z}^{z_0} {d\zeta \over \zeta^3}\frac{J_{0i}}{h} \right]
       -\frac{1}{3} i k_i F\left[\int_z^{z_0} {d\zeta \over
         \zeta^3} J_{j}^{j}\right]
       -J_{i5}h\zeta^{-3}\big|_{z_0}+J_{i5}^{(3)}\,.
\end{align}
Note that the last two terms cancel each other at order $\mathcal{O}(k^{-1}J_{m5}^{(3)})$. This follows from expanding \eqref{E:DJeq0Jm5} in small $k$ and solving for $z$. To emphasize this point we substitute $J_{i5}^{(3)}$ with $\Delta j_{i5}^{(3,\overline{1})}$.
Next, we note that generically, $J_{0i}h^{-1}$ diverges near the horizon. Since $\mathcal{F}$ is finite, this divergence must be canceled by a similar divergence which can only come from the next to last term. Thus, we write
\begin{equation}
	i k_j \mathcal{F}_{ij}/\Delta = -i\omega{\rm F} \left[- i \frac{j_{i5}^{(\overline{1})} h}{\omega\zeta^3} \Big|_{z_0} + \int_{z}^{z_0} {d\zeta \over \zeta^3} \frac{j_{0i}^{(\overline{0})}}{h} \right]
       -\frac{1}{3} i k_i {\rm F} \left[\int_{z}^{z_0} {d\zeta \over \zeta^3} \, j^{j\,(\overline{0})}_{j} \right]
       +j_{i5}^{(3,\overline{1})}+\mathcal{O}(k^2)\,.
\end{equation}
If
\begin{equation}
\label{E:NoWakeCondition}
	\lim_{z \to z_0} J_{i5}^{(\overline{1})}h = 0\,,
\end{equation}
then the integral
\begin{equation}
	\int_z^{z_0} \frac{j_{0i}^{(\overline{1})}\zeta^{-3}}{h }\Delta d\zeta
\end{equation}
must be finite for $z>0$ and
the $1/\omega$ pole will be canceled by the $\omega$ dependence of the first term on the right hand side of \eqref{E:SD0}. We end up with
\begin{align}
    \epsilon & =
        i\frac{3\omega}{\vec{k}^2-3\omega^2}J_{05}^{(3)}
		+\frac{\vec{k}^2}{\vec{k}^2-3\omega^2}{\rm F}\left[\int_z^{z_0}{d\zeta \over \zeta^3} J^{l}_{l}\right]
    		+\frac{3\omega k_j}{\vec{k}^2-3\omega^2}{\rm F}\left[\int_z^{z_0}{d\zeta \over \zeta^3} J_{0j}h^{-1}\right]\\
    S_i &=i\frac{k_i}{\vec{k}^2-3\omega^2}J_{05}^{(3)}+\frac{k_i\omega}{\vec{k}^2-3\omega^2}
        {\rm F}\left[\int_z^{z_0}{d\zeta \over \zeta^3} J^{l}_{l}\right]
    +\frac{k_i k_j - \delta_{ij}(\vec{k}^2-3\omega^2)}{\vec{k}^2-3\omega^2}
    {\rm F}\left[\int_z^{z_0}{d\zeta \over \zeta^3} J_{0j}h^{-1}\right]\,.
\end{align}
Since the $\omega^{-1}$ pole has been canceled, condition \eqref{E:NoWakeCondition} is sufficient for the absence of a wake as long as $\Delta$ does not introduce such a pole. Comoving string configurations which do not reach the horizon and whose monopole structure of $j_{m5}^{(3)}$ vanishes (for instance mesons or baryonic configurations) fall into this category.

\section{Short distance asymptotics}
\label{S:Short}
In the previous section we extracted the large distance asymptotics of $H_{mn}^{(4)}$ by appealing to a small momentum approximation, $k z_0 \ll 1$. Here, we consider the short distance asymptotics of the solution, associated with large momentum: $k z_0 \gg 1$. The $k z_0 \to \infty$ limit corresponds to taking the temperature of the black hole to zero; as long as we keep terms up to order $\mathcal{O}(k^{-4})$, we are essentially working in the AdS geometry whose line element is given by \eqref{E:AdSBH} but with $h=1$.
More explicitly, by choosing
\begin{equation}
\label{E:DefX}
	X = \left(H_{(11)},H_{(12)},H_{(13)},H_{(22)},H_{(23)},-\frac{3}{h}H_{00}+H_{i}^{i},H_{i}^{i},H_{01},H_{02},H_{03}\right),
\end{equation}
the second order linearized Einstein equations \eqref{E:DHeqJ} may be written as
\begin{equation}
\label{E:XEOM}
	\alpha^{-3}h^{-n_i}\partial_z \left(X_i^{\prime} \alpha^3 h^{n_i} \right)
	+V_{ij} X_j = -\frac{4}{h}S_i
\end{equation}
where
\begin{align}
\label{E:ni}
	n &= \left(1,1,1,1,1,3/2,1/2,0,0,0\right)\\
\label{E:Si}
    S &= \left(J_{(11)},J_{(12)},J_{(13)},J_{(22)},J_{(23)},
        -\left(\frac{1}{h}J_{00}+J_{ii}\right),\frac{1}{h}J_{00},J_{01},J_{02},J_{03}\right)
\end{align}
and $V_{ij}$ is an $\mathcal{O}(k^2)$ $10 \times 10$ matrix given in appendix \ref{A:EOM}. In the $k z_0 \gg 1$ approximation $h=1+\mathcal{O}(z_0^{-4})$, so up to order $(z_0 k)^{-4}$ all the kinetic terms in \eqref{E:XEOM} have the same form and $V_{ij}$ is a constant matrix. The left hand side of \eqref{E:XEOM} reduces to the equations of motion for linearized gravity in an AdS background and the source terms on the right hand side of \eqref{E:XEOM} need to be expanded up to order $(z_0 k)^{-4}$.\footnote{Higher order corrections coming from the source term may also be treated. This is discussed in \cite{Gubser:2007nd}. See also \cite{Gubser:2007xz} for an explicit application.}

To solve \eqref{E:XEOM} in the $k z_0 \gg 1$ approximation, consider the five eigenvectors,
$\chi^i$, of $V_{ij}^T$,
whose eigenvalues are $-k^2 = \omega^2-\vec{k}^2$, described in appendix \ref{A:emptyAdS5}. From \eqref{E:XEOM}, the combinations $Y^i = \chi^i_{j} X_j$ satisfy the massless scalar field equation in empty AdS space
\begin{equation}
\label{E:EOMY}
    \left(\alpha^{-3}\partial_z \alpha^3 \partial_z -k^2 \right)Y^i = -4 \chi^i_{j} S_j\,.
\end{equation}

The Green's function for the operator on the left hand side of \eqref{E:EOMY} has been
extensively studied in the literature, see for example \cite{Danielsson:1998wt}. Using the Green's function, the solution to \eqref{E:EOMY} may be written as an integral over the source term \cite{Lin:2007fa}. We follow a somewhat different path to solve \eqref{E:EOMY}, which allows us to write the solution to the equations of motion as a sum instead of an integral. 
Let's assume that
 \eqn{JexpressHighK}{
  J_{\mu\nu}(\omega,\vec{k},z) = \Delta^{\hbox{c}}(\omega,\vec{k})
    j_{\mu\nu}^{\hbox{c}}(\omega,\vec{k},z) \,.
 }
where $j_{\mu\nu}^{\hbox{c}}(\omega,\vec{k},z)$ has an expansion in integer powers of $\omega$ and $\vec{k}$ for large values of $\omega$ and $\vec{k}$, whereas $\Delta^{\hbox{c}}(\omega,\vec{k})$ can be arbitrarily non-analytic.
We expand 
 \eqn{sjaExpand}{
  S_j = \sum_{a=-1}^\infty S_j^{(a)} z^a + {\cal O}(k^{-4}\Delta^{\hbox{c}}) \,.
 }
According to the discussion at the beginning of this section, we are then allowed to solve \eno{E:XEOM} with $S_i$ replaced 
by the right hand side of \eno{sjaExpand} and with $h=1$. 
In appendix \ref{A:emptyAdS5} we construct the solution $W_n(z)$ to the equation
\begin{equation}
	\left(\alpha^{-3}\partial_z \alpha^3 \partial_z - k^2 \right)W_n(z) = z^n
\end{equation}
with appropriate boundary conditions for any $n\geq -1$.
Thus, the solution to \eno{E:EOMY} takes the form
\begin{equation}
	Y^i = -4 \chi^{i}_j\sum_{a=-1}^{\infty} W_a(z) S_j^{(a)}+\mathcal{O}(k^{-4}\Delta^{\hbox{c}})\,.
\end{equation}

Recall that we are not interested in the full solution to \eqref{E:EOMY}, but only in the fourth order coefficient of $Y^{i}$:
\begin{equation}
\label{E:Yi4}
	Y^{i\,(4)}=-4 \chi^i_{j} \sum_{a=-1}^{\infty}W_a^{(4)} S_j^{(a)}+\mathcal{O}(k^{-4}\Delta^{\hbox{c}})\,.
\end{equation}
Using \eqref{E:Specialw}, \eqref{E:wtowBessel}, \eqref{E:bvals} and \eqref{E:avals} we find
\begin{equation}
	W_n^{(4)}(ik)^{n-2} = \begin{cases}
		     -\frac{3}{16}+\frac{1}{4}\gamma_E & n=2 \\
		     -\frac{3}{64}+\frac{1}{16}\gamma_E & n=0 \\
	             (-1)^{\frac{n}{2}}2^{n-5}\Gamma\left(\frac{1}{2}n+1\right)\Gamma\left(\frac{1}{2}n-1\right) & \hbox{otherwise,}
	            \end{cases}
\end{equation}
where $\gamma_E$ is the Euler-Mascheroni constant.

With $Y^{i\,(4)}$ in hand we can solve $\chi^i_j X_{j}^{(4)}=Y^{i\,(4)}$, the energy conservation condition \eqref{E:Constraint}, and the trace equation \eqref{E:Trace} to get the energy momentum tensor in terms of $J_{m5}^{(3)}$ and $Y^{i\,(4)}$. Defining
\begin{equation}
\label{E:DefJJmn}
	\mathcal{F}_{mn}^{\hbox{c}}  = -4 \sum_{a=-1}^{\infty} W_a^{(4)} J_{mn}^{(a)}+\mathcal{O}(k^{-4}\Delta^{\hbox{c}}),
\end{equation}
and
\begin{align}
	F_m^{\hbox{c}} & = J_{m5}^{(3)} -i k^m \mathcal{F}^{\hbox{c}}_{mn}\\
		& = J_{m5}^{(3)} + 4 \sum_{a=0}^{\infty} W_{a-1}^{(4)} (3-a) J_{n5}^{(a)}
\end{align}
(where in the last line we made use of \eqref{E:DJeq0Jm5}, which is valid to all orders in $a$ in the $\mathcal{O}(k^{-4})$ approximation we are using) and making use of \eqref{deltaTmn}, the fluctuations of the stress energy tensor take the form
\begin{equation}
\label{E:TmnShort}
	\langle \delta T_{mn} \rangle = T_{mn}^{\hbox{c}} + \mathcal{F}^{\hbox{c}}_{mn}+\mathcal{A}_{mn}^{\hbox{c}}
\end{equation}
with
\begin{equation}
\label{E:TmnC}
	T^{\hbox{c}}_{mn} =
	\frac{i}{3k^4} k^l F_l^{\hbox{c}} \left(2 k_m k_n +\eta_{mn} k^2\right)
	-\frac{i}{k^2}\left(k_m F_n^{\hbox{c}}+k_n F_m^{\hbox{c}}\right)
	+\frac{1}{3k^2}\mathcal{F}^{\hbox{c}\,l}_{l}\left(k_mk_n-\eta_{mn}k^2\right).
\end{equation}
and 
\begin{equation}
\label{E:Tcanomaly}
	\mathcal{A}_{mn}^{\hbox{c}} =
	\frac{18}{24}\frac{L^3}{\kappa_5^2}R^{(2)}_{mn}
	+\frac{1}{24}\frac{1}{k^2}\left(5\eta_{mn}k^2+4 k_m k_n\right)J_{m}^{(2)\,m}-\frac{1}{4}J_{mn}^{(2)}
	+\frac{1}{k^2}\left(k_m k_n - \eta_{mn} k^2\right)\langle \delta T^l_l \rangle
\end{equation}
where we have made use of \eqref{E:R2properties}. The trace of $\mathcal{A}_{mn}$, $\langle \delta T_l^l \rangle$, is given in \eqref{E:TraceT}.
Note that
\begin{equation}
	i k^m T_{mn}^{\hbox{c}} = F^{\hbox{c}}_n.
\end{equation}

\section{Strings and point particle sources}
\label{S:Examples}
In this section we consider two specific sources: stringy sources whose motion may be determined through the Nambu-Goto action
\begin{equation}
\label{E:NG}
	\mathcal{S}_{NG} = \int d\tau d\sigma \, \sqrt{-g} \, \mathcal{L}_{NG}\quad\qquad\mathcal{L}_{NG} = -\frac{1}{2\pi\alpha^{\prime}}\sqrt{-g}
\end{equation}
with $g_{\alpha\beta} = \partial_{\alpha}X^{\mu}\partial_{\beta}X^{\nu}G_{\mu\nu}$,
and point particles with mass $m \geq 0$ whose action is given by
\begin{equation}
\label{E:Sparticle}
	S_{\hbox{\tiny particle}} = \int d\eta \, L_{\hbox{\tiny particle}} \quad\qquad L_{\hbox{\tiny particle}}= \frac{1}{2e}G_{\mu\nu}\dot{X}^{\mu}\dot{X}^{\nu}-\frac{1}{2}m^2e \,,
\end{equation}
where $e$ is a Lagrange multiplier. In both instances the four-force induced by these objects in the boundary theory can be calculated for generic configurations. We find that a drag force acts on the string endpoint if it is located on the asymptotically AdS boundary while point particles, which are localized on the boundary only at some time $t=T_{(b)}$, supply an impulse to the plasma at that time.

\subsection{Strings}
\label{SS:Strings}
Strings whose endpoints are on the asymptotically AdS boundary are dual to infinitely massive quarks \cite{Herzog:2006gh,Gubser:2006bz}, mesons \cite{Maldacena:1998im,Chernicoff:2006hi}, or baryons if one introduces D5-brane baryon vertices \cite{Witten:1998xy,Gross:1998gk}.

The energy momentum tensor \eqref{SMvary} following from the Nambu-Goto Action \eqref{E:NG}
is given by
\begin{equation}
	J^{mn}(t,\vec{x},z) = - \frac{L^3}{2\pi\alpha^{\prime}}\int d\sigma \, \sqrt{-g} \, \alpha^{-5} \delta^{(3)}(\vec{x}-\vec{X}(t,\sigma))\delta(z-Z(t,\sigma)) \partial_{\alpha}X^{\mu}\partial^{\alpha}X^{\nu} 
\end{equation}
where we have used a gauge where $t=\tau$ so that $X^{\mu} = \begin{pmatrix} t & X^1 & X^2 & X^3 & Z \end{pmatrix}$. We'll be interested in the boundary four-force density generated by the string. Therefore, we focus on the near boundary behavior of $J_{m5}$ where we can, at least locally, use the gauge $Z=\sigma$. A near boundary expansion gives us
\begin{align}
	J_{05} &=  \frac{L^3}{2\pi\alpha^{\prime}} \alpha^{-1} \delta^{(3)}(\vec{x}-\vec{X}(t,z)) \frac{g_{10}}{\sqrt{-g}} +\ldots\\
	J_{i5} &= \frac{L^3}{2\pi\alpha^{\prime}} \alpha^{-1} h^{-1} \delta^{(3)}(\vec{x}-\vec{X}(t,z)) \frac{\left(g_{00} {X}^{i\,\prime}(t,z) - g_{10} \dot{X}^{i}(t,z)\right)}{\sqrt{-g}}+\ldots.
\end{align}
Defining $P_i$ as the $\sigma$ component of the world-sheet current,
\begin{equation}
\label{E:DefPi}
	P_i \equiv \frac{\delta S_{NG}}{\delta X^{i\,\prime}} =
	\frac{1}{2 \pi\alpha^{\prime}} \frac{\alpha^2}{\sqrt{-g}} \left(g_{00}X^{i\,\prime} - g_{01}\dot{X}^{i}\right)
\end{equation}
(which is not necessarily a constant)
we find that near the boundary,
\begin{align}
	J_{05}^{\rm string} &= - z^3 \delta^{(3)}(\vec{x}-\vec{X}(t,z)) \dot{X}^i P_i +\ldots\\
	J_{i5}^{\rm string} &= z^3 \delta^{(3)}(\vec{x}-\vec{X}(t,z)) P_i + \ldots.
\end{align}
Using \eqref{E:Tmnsourced} and assuming a finite $P_i$ on the boundary, we find that the four force is localized at the string endpoint(s) $X^{i} = X^{i}_{(b)}$ and is given by
\begin{equation}
 	J_{m5}^{(3)} =
	\sum_b \delta^{(3)}(\vec{x}-\vec{X}_{(b)})\lim_{z \to 0}
				     \begin{pmatrix}
	                             -\vec{V}\cdot \vec{P} &
				     \vec{P}
	                             \end{pmatrix}
\end{equation}
where $\vec{V}$ is the velocity of the string endpoint, $\vec{V} = \lim_{z\to0}\dot{\vec{X}}$, and the sum is over the endpoints of the string which reach the boundary. A trailing string will clearly induce a wake since the total drag force is non vanishing. The total drag force acting on the center of mass of mesonic and baryonic configurations (which were recently introduced in this context in \cite{Athanasiou:2008pz}) vanishes and therefore, from \eqref{E:Sparticle} and the condition \eqref{E:NoWakeCondition} a wake will not form (at least to leading order.)

In order to obtain the $\mathcal{F}_{mn}$ corrections, we need to consider an explicit string configuration. The large distance and short distance asymptotics of $\langle \delta T_{mn} \rangle$ for various string configurations has been studied in \cite{Gubser:2007ga, Gubser:2007xz, Gubser:2007ni, Chesler:2007sv, Gubser:2007zr, Friess:2006fk, Chesler:2007an, Gubser:2007nd, Yarom:2007ni}. We have checked that our main result \eqref{MainResult} fits with the expressions in the literature. The reader is referred to these references for the detailed structure of the stress-energy tensor.

In sections \ref{S:EMTensor} and Appendix \ref{A:HoloRG} we argued that the stress-energy tensor contains divergent terms which are localized to the extent that the probe-source is localized. For the case of string-configurations we will now show explicitly that these divergences can be associated with the infinite mass of the quark dual to the string endpoint. Since $P_i$ and the velocity $V_i$ are finite at the asymptotically AdS boundary, we find from \eqref{E:DefPi} that near the boundary, in the $Z=\sigma$ gauge, $X_i^{\prime}(t,z)$ is at least of order $z^2$. This implies that
\begin{align}
\label{E:NearBString}
    g_{00} &= -\alpha^2 \left(1-V^2\right)+\mathcal{O}(z^0)\\
    g_{10} &= \mathcal{O}(z^0)\\
    g_{11} &=\alpha^2+\mathcal{O}(z^0).
\end{align}
Thus, to leading order in $z$, we find that a string endpoint which reaches the AdS boundary will induce an $\mathcal{O}(z)$ contribution to $J_{mn}$ of the form
\begin{equation}
\label{E:Jmn1}
    J_{mn}^{(1)} = M_s \epsilon \sqrt{1-V^2} U_m U_n \delta^{(3)}(\vec{x}-\vec{X}_{(b)})
\end{equation}
where
\begin{equation}
    U_m = \begin{pmatrix} -1 & V_i \end{pmatrix}/\sqrt{1-V^2}
\end{equation}
is the four velocity of the quark
and
\begin{equation}
    M_s = \frac{L^2}{2\pi\alpha^{\prime}\epsilon}
\end{equation}
is the mass of a static quark dual to a string ending on a D7-flavor brane \cite{Karch:2002sh} a distance $z=\epsilon$ from the boundary \cite{Herzog:2006gh}. In the $\epsilon \to 0$ limit, or in the absence of D7-branes, the mass of the quark becomes infinite.
With \eqref{E:Jmn1}, we find that the divergent contribution of the stringy source to the stress energy tensor takes the form:
\begin{equation}
\label{E:MassContribution}
    \langle T_{mn}^{\epsilon} \rangle = \sum_b M_s \sqrt{1-V^2} U_m U_n \delta^{(3)}(\vec{x}-\vec{X}_{(b)})
\end{equation}
which is precisely what we would expect from a quark of (infinite) mass $M_s$ moving with a 4-velocity $U^{m}$. Naturally, it is localized at the position of the quark. Our result \eqref{E:MassContribution} is an extension of the one obtained in \cite{Chesler:2007sv} for the special case of a trailing string moving at constant velocity. In the case of a stringy source there are no other divergent contributions to the stress-energy tensor.

\subsection{Point particles}
\label{SS:Pointparticles}
Here we discuss in some detail pointlike particles in the AdS black hole background. These can be thought of as approximations to quantum states of closed strings propagating in the bulk, which represent color-singlet quasi-particle excitations of the gauge theory---heuristically, glueballs which are in the process of thermalizing with the medium.
Before computing the stress tensor, let us introduce
\begin{equation}
\label{E:particlemomentum}
	P_m = \frac{\partial L}{\partial \dot{X}^m} = \frac{1}{e}G_{m\nu}\dot{X}^{\nu}
\end{equation}
which is conserved since $G_{\mu\nu}$ is independent of $X^m$. Thus, we are free to define
\begin{equation}
	P_m = \begin{pmatrix} -E & P_1 & P_2 & P_3 \end{pmatrix}
\end{equation}
with $E$ and $P_i$ constants.
The stress tensor of the point particle following from varying the action \eqref{E:Sparticle} with the conventions we introduced in \eqref{SMvary} is
\begin{equation}
\label{E:Jmnparticle}
	J^{mn} = \frac{L^3}{\sqrt{-G}}\delta^{(3)}(x-X(t))\delta(z-Z(t)) \frac{1}{e}\dot{X}^{\mu}\dot{X}^{\nu}
\end{equation}
where we have used the gauge $X^{\mu} = \begin{pmatrix} t & X^1 & X^2 & X^3 & Z \end{pmatrix}$. Using \eqref{E:particlemomentum} to evaluate the near boundary asymptotics of $J_{m5}$ we find
\begin{align}
	J_{m5} &= z^3 \delta^{(3)}(x-X(t))\delta(z-Z(t)) P_m \dot{Z}+\ldots\\
	&=z^3 \delta^{(3)}(x-X(z))\delta(t-T(z))P_m+\ldots,
\end{align}
so that
\begin{equation}
	J_{m5}^{(3)} = \delta(t-T_{(b)})\delta^{(3)}(x-\vec{X}_{(b)}) P_m
\end{equation}
where $T_{(b)}$ and $X_{(b)}$ are the time and place where the particle reaches the boundary.

To study the stress tensor of the plasma in response to the glueball we need to restrict ourselves to a more specific particle configuration. Consider a massless particle traveling along $x^2 = x^3 = 0$ with $E>P_1>0$. Using the conservation equations \eqref{E:particlemomentum} and the equation of motion for $e$ which follows from the variation of \eqref{E:Sparticle}, one finds that the trajectory of such a particle satisfies
\begin{align}
\label{E:dotX1}
	\dot{X}^1 & = \frac{P_1}{E}h(Z)\\
\label{E:dotZ}
	\dot{Z} &= h(Z) \sqrt{1-\frac{P_1^2}{E^2}h(Z)}
\end{align}
where we have chosen a solution representing a point particle moving from the boundary to the horizon. Setting $X^1(0)=Z(0)=0$ we get
\begin{equation}
\label{E:Jm53particle}
 	J_{m5}^{(3)} = \delta(t)\delta^{(3)}(\vec{x})\begin{pmatrix} -E & P_1 & 0 & 0 \end{pmatrix}.
\end{equation}
According to our discussion in section \ref{SS:Naive} this is enough to determine the leading long distance components of the stress tensor and the subleading inviscid corrections to it. The non-hydrodynamical subleading corrections manifest themselves in terms of $\mathcal{F}_{mn}$, defined in \eqref{FResult}. From \eqref{E:Jmnparticle}, \eqref{E:dotX1} and \eqref{E:dotZ} we find
\begin{equation}
 	J_{(ij)} = \frac{P_1^2 z^3 h(z)}{E} e^{-i\omega T(z)+i k_1 X^1(z)}
		\begin{pmatrix}
		 	-\frac{2}{3} & 0 & 0   \\
			0 & \frac{1}{3} & 0  \\
			0 & 0 & \frac{1}{3}
		\end{pmatrix}
\end{equation}
where $T(Z(t))=1$. It follows that
\begin{equation}
\label{E:Fparticle}
 	\mathcal{F}_{ij} = \frac{P_1^2 z_0^4}{8 E}
			\begin{pmatrix}
		 	-\frac{2}{3} & 0 & 0   \\
			0 & \frac{1}{3} & 0  \\
			0 & 0 & \frac{1}{3}
		\end{pmatrix} \,.
\end{equation}

The energy momentum tensor of this configuration can be read off of \eqref{MainResult} by using \eqref{THydro}, \eqref{E:resummed}, \eqref{E:Jm53particle} and \eqref{E:Fparticle}. The resulting expressions are somewhat long as are their counterparts in real space. To get a flavor for the dynamics of the decaying glueball, consider a simplified trajectory where $P_1=0$: a massless particle located at the boundary at $t=0$ and moving straight down into $AdS_5$-Schwarzschild with $x^1=x^2=x^3=0$. In this case, the leading term for the energy density reads
\begin{equation}
	\epsilon(\omega,\vec{k}) = -\frac{3i\omega E}{\vec{k}^2-3\omega^2-3 i \Gamma_s \vec{k}^2\omega} \,,
\end{equation}
where in the denominator we have added viscous corrections, $\Gamma_s = \frac{1}{3\pi T}$, which enable us to carry out the Fourier transform in the $\omega$ coordinate via contour integration. When  Fourier transforming in the momentum coordinates we keep $k$ small but allow $k t$ to be large \cite{Casalderrey-Solana:2004qm}. We find
\begin{equation}
\label{E:EnergyinBH}
    \epsilon(t,\vec{x}) =
    -\frac{E}{4\pi r}\partial_t \frac{\left(e^{-\frac{(r+c_s t)^2}{2\Gamma_s t}}-e^{-\frac{(r-c_s t)^2}{2\Gamma_s t}}\right)}{\sqrt{2 \Gamma_s t}}\Theta(t).
\end{equation}
where $\Theta(t)$ is the step function which vanishes for $t<0$ and $r^2 = x^i x^i$.
Up to the step function, equation \eqref{E:EnergyinBH} coincides with the kernel for the sound mode contribution to the energy in the linear hydrodynamic approximation which was calculated in \cite{Casalderrey-Solana:2004qm}. This is not surprising since the point particle provides an instantaneous impulse to the plasma.

For the leading short distance asymptotics of the stress tensor, we need to compute $\mathcal{F}^{\hbox{c}}_{mn}$ defined in \eqref{E:DefJJmn}. Using $\eta = \sqrt{E^2-P_1^2}$, we find that the leading contribution to $\mathcal{F}^{\hbox{c}}_{mn}$ reads
\begin{multline}
 	\mathcal{F}_{mn}^{\hbox{c}} =
	-4\frac{E}{\eta}\begin{pmatrix}
	 	E & - P_1 & 0 & 0 & -\eta \\
		-P_1 & \frac{P_1^2}{E} & 0 & 0 & \frac{P_1}{E}\eta \\
		0 & 0 & 0 & 0 & 0 \\
		0 & 0 & 0 & 0 & 0 \\
		-\eta & \frac{P_1}{E}\eta & 0 & 0 & 1-\frac{P_1^2}{E}
	\end{pmatrix}
	\times
	\\ \sum_{a=3}^{\infty}(-1)^{a/2}2^{a-5}\Gamma\left(\frac{1}{2}a+1\right)\Gamma\left(\frac{1}{2}a-1\right)q^{2-a}
	\frac{\left(i\frac{\omega E -k_1 P_1}{\eta}\right)^{a-3}}{(a-3)!}.
\end{multline}
The sum on the right hand side can be evaluated explicitly:
\begin{equation}
\label{E:FmnCparticle}
 	\mathcal{F}_{mn}^{\hbox{c}} =
	-4\left(
		\frac{(5+2 \xi^2)\xi}{8 q(1+\xi^2)^2}
		+\frac{3 \sinh^{-1}(\xi)-\frac{1}{2}i\pi}{8 q(1+\xi^2)^{5/2}}
	\right)
	\frac{E}{\eta}\begin{pmatrix}
	 	E & - P_1 & 0 & 0 & -\eta \\
		-P_1 & \frac{P_1^2}{E} & 0 & 0 & \frac{P_1}{E}\eta \\
		0 & 0 & 0 & 0 & 0 \\
		0 & 0 & 0 & 0 & 0 \\
		-\eta & \frac{P_1}{E}\eta & 0 & 0 & 1-\frac{P_1^2}{E}
	\end{pmatrix}
\end{equation}
where $\xi = \frac{\omega E - k_1 P_1}{\eta k}$ and $k^2 = -\omega^2+\vec{k}^2$.
By plugging \eqref{E:FmnCparticle} into \eqref{E:TmnShort} we obtain the leading short distance asymptotics of the stress tensor.
The expressions we find are somewhat complicated, so we will not reproduce them here.  Instead, we pass to a special case: we consider once again the $P_1 \to 0$ limit and assume that $\omega$ has a slightly positive imaginary part.
Then,
\begin{equation}
\label{E:Eparticleshort}
	\epsilon(\omega,\vec{k}) = i \frac{E}{|\vec{k}|}\tanh^{-1}\left(\frac{|\vec{k}|}{\omega}\right).
\end{equation}
Fourier transforming \eqref{E:Eparticleshort} we find
\begin{equation}
\label{E:EnergyinAdS}
 	\epsilon(t,\vec{x}) = \frac{E}{4\pi r^2}\delta(t-r)
\end{equation}
where we have used $\hbox{Im}\{\omega\}>0$. We note that this expression can be obtained in a rather straightforward way by appealing to the symmetries of the problem. Using energy conservation, the ${O}(3)$ symmetry in the $x^1,x^2,x^3$ directions, causality, and the fact that the point-like particle moving in the AdS bulk perturbs only a light-like component of the metric \cite{Aichelburg:1970dh}, we find that
\begin{equation}
\label{E:Energyfromsymmetries}
    \langle \delta T_{uu} \rangle = \frac{E}{4\pi r^2}\delta(u)
\end{equation}
with $u = t-r$.

From \eqref{E:EnergyinAdS} and~\eqref{E:EnergyinBH} we see that the energy density of a glueball which is ``injected'' into the plasma at $t=0$ will initially propagate outward with the speed of light, but at large distances, where hydrodynamics kick in, it will propagate with the speed of sound.
We note that \eqref{E:EnergyinAdS} is in some disagreement with a result obtained in \cite{Lin:2007fa}, where it was claimed that the boundary stress energy tensor related to a massive particle moving in empty AdS vanishes. 

\section*{Acknowledgments}

We thank S.~Pufu for collaboration on the early stages of this project.  S.S.G.~thanks G.-L. Ma for useful discussions.  A.Y.~would like to thank M. Haack and R. Helling for useful discussions.  The work of S.S.G.\ was supported in part by the Department of Energy under Grant No.\ DE-FG02-91ER40671 and by the NSF under award number PHY-0652782.  A.Y. is supported in part by the Minerva foundation and by the German science foundation.

\begin{appendix}
\section{Linearized Einstein equations in the AdS${}_5$-Schwarz\-schild background}
\label{A:EOM}
The second order Einstein equations \eqref{E:DHeqJ} for the linearized metric fluctuations in an AdS${}_5$ black hole background \eqref{E:Gmn} are given by
\begin{equation}
\label{E:EOMforX}
	\alpha^{-3}h^{-n_i}\partial_z \left(X_i^{\prime} \alpha^3 h^{n_i} \right)
	+V_{ij} X_j = -\frac{4}{h}S_i
\end{equation}
where $X_i$, $n_i$ and $S_i$ are defined in \eqref{E:DefX}, \eqref{E:ni} and \eqref{E:Si}
and $V_{ij}$ is an $\mathcal{O}(k^2)$ $10 \times 10$ matrix
\begin{equation}
    V = V_0 + V_1 + V_2
\end{equation}
with
\begin{equation}
	V_2 =
	\begin{pmatrix}
		\frac{\omega^2}{h^2} & 0 & 0 & 0 & 0 & 0 & 0 & 0 & 0 & 0\\
		0 & \frac{\omega^2}{h^2} & 0 & 0 & 0 &0 & 0 & 0 & 0 & 0 \\
		0 & 0 & \frac{\omega^2}{h^2} & 0 & 0 & 0 & 0 & 0 & 0 & 0 \\
		0 & 0 & 0 & \frac{\omega^2}{h^2} & 0 & 0 & 0 & 0 & 0 & 0 \\
		0 & 0 & 0 & 0 & \frac{\omega^2}{h^2} & 0 & 0 & 0 & 0 & 0 \\
		0 & 0 & 0 & 0 & 0 & 0 & \frac{2\omega^2}{h^2} & 0 & 0 & 0 \\
		0 & 0 &	0 & 0 & 0 & 0 & 0 & 0 & 0 & 0 \\
		0 & 0 &	0 & 0 & 0 & 0 & 0 & 0 & 0 & 0 \\
		0 & 0 &	0 & 0 & 0 & 0 & 0 & 0 & 0 & 0 \\
		0 & 0 &	0 & 0 & 0 & 0 & 0 & 0 & 0 & 0 \\
	\end{pmatrix}
\end{equation}
\begin{equation}
	V_1 =
	\begin{pmatrix}
    0 & 0 & 0 & 0 & 0 & 0 & 0 & \frac{4\,k_1\,\omega }{3\,h^2} & \frac{-2\,k_2\,\omega }{3\,h^2} &
     \frac{-2\,k_3\,\omega }{3\,h^2} \\
    0 & 0 & 0 & 0 & 0 & 0 & 0 & \frac{k_2\,\omega }{h^2} & \frac{k_1\,\omega }{h^2} & 0 \\
    0 & 0 & 0 & 0 & 0 & 0 & 0 & \frac{k_3\,\omega }{h^2} & 0 & \frac{k_1\,\omega }{h^2} \\
    0 & 0 & 0 & 0 & 0 & 0 & 0 & \frac{-2\,k_1\,\omega }{3\,h^2} &
     \frac{4\,k_2\,\omega }{3\,h^2} & \frac{-2\,k_3\,\omega }{3\,h^2} \\
    0 & 0 & 0 & 0 & 0 & 0 & 0 & 0 & \frac{k_3\,\omega }{h^2} & \frac{k_2\,\omega }{h^2} \\
    0 & 0 & 0 & 0 & 0 & 0 & 0 & \frac{4\,k_1\,\omega }{h^2} & \frac{4\,k_2\,\omega }
     {h^2} & \frac{4\,k_3\,\omega }{h^2} \\
    0 & 0 & 0 & 0 & 0 & 0 & 0 & 0 & 0 & 0 \\
    -\frac{k_1\,\omega }{h}   & -\frac{k_2\,\omega }{h}  &
     -\frac{k_3\,\omega }{h}  & 0 & 0 & 0 & \frac{2\,k_1\,\omega }
     {3\,h} & 0 & 0 & 0 \\
    0 & -\frac{k_1\,\omega }{h}   & 0 & -\frac{k_2\,\omega }{h}   & -\frac{
     k_3\,\omega }{h}   & 0 & \frac{2\,k_2\,\omega }{3\,h} & 0 & 0 & 0 \\
    \frac{k_3\,\omega }{h} & 0 & - \frac{k_1\,\omega }{h}   & \frac{k_3\,\omega }{h} & -
     \frac{k_2\,\omega }{h}   & 0 & \frac{2\,
     k_3\,\omega }{3\,h} & 0 & 0 & 0
    \end{pmatrix}
\end{equation}
\begin{multline}
    V_0 =\frac{1}{h}
    \hbox{\tiny $\begin{pmatrix}
    \frac{1}{3}{k_1}^2 - {{k_2}^2} - \frac{1}{3}{k_3}^2 & \frac{2}{3}k_1k_2 &
    \frac{2}{3}k_1k_3 & -\frac{2}{3}\left( {k_2}^2 - {k_3}^2 \right) & -\frac{4}{3}k_2
     k_3 & -\frac{2}{9}k_1^2+\frac{1}{9}k_2^2+\frac{1}{9}k_3^2 & 0 & 0 & 0 & 0 \\
     {k_1k_2}
    & -{{k_3}^2}  & {k_2 k_3} & {k_1 k_2} &
     k_1 k_3 & -\frac{1}{3} k_1\,k_2   & 0 & 0 & 0 & 0 \\
     0 & {k_2 k_3} & -{{k_2}^2}   & -{k_1 k_3}  & {k_1 k_2}
    & -\frac{1}{3} k_1 k_3  & 0 & 0 & 0 & 0 \\
   -\frac{2}{3}\left( {k_1}^2 - {k_3}^2 \right)  & \frac{2}{3} k_1 k_2  & -\frac{4}{3} k_1 k_3  & -{{k_1}^2}  +
   \frac{1}{3}{k_2}^2 - \frac{1}{3}{k_3}^2 & \frac{2}{3} k_2 k_3  & \frac{1}{9}\left({k_1}^2 -
     2 {k_2}^2 + {k_3}^2\right) & 0 & 0 & 0 & 0 \\
    -{k_2 k_3}  & {k_1
     k_3} & {k_1 k_2} & 0 & -{{k_1}^2}  & -\frac{1}{3} k_2
       k_3   & 0 & 0 & 0 & 0 \\
       0 & 0 & 0 & 0 & 0 & -\frac{2}{3}
     k^2  & \frac{2}{3} k^2  & 0 & 0 & 0 \\
   {{k_1}^2 - {k_3}^2} & {2
     k_1 k_2} & {2 k_1 k_3} & {{k_2}^2 - {k_3}^2} & {2
     k_2 k_3} & 0 & -\frac{2}{3} k^2
     & 0 & 0 & 0 \\
   0 & 0 & 0 & 0 & 0 & 0 & 0 & 0   & 0
    & 0 \\
   0 & 0 & 0 & 0 & 0 & 0 & 0 & 0   & 0
    & 0 \\
   0 & 0 & 0 & 0 & 0 & 0 & 0 & 0   & 0
    & 0
    \end{pmatrix}$}
\\+
	\frac{1}{h}\begin{pmatrix}
   0 & 0 & 0 & 0 & 0 & 0 & 0 & 0   & 0
    & 0 \\
   0 & 0 & 0 & 0 & 0 & 0 & 0 & 0   & 0
    & 0 \\
   0 & 0 & 0 & 0 & 0 & 0 & 0 & 0   & 0
    & 0 \\
   0 & 0 & 0 & 0 & 0 & 0 & 0 & 0   & 0
    & 0 \\
   0 & 0 & 0 & 0 & 0 & 0 & 0 & 0   & 0
    & 0 \\
   0 & 0 & 0 & 0 & 0 & 0 & 0 & 0   & 0
    & 0 \\
   0 & 0 & 0 & 0 & 0 & 0 & 0 & 0   & 0
    & 0 \\
   0 & 0 & 0 & 0 & 0 & 0 & 0 & - {k_2}^2 - {k_3}^2   & {k_1 k_2}
    & k_1 k_3 \\
   0 & 0 & 0 & 0 & 0 & 0 & 0 & {k_1 k_2} & -{k_1}^2 -
       {k_3}^2  & {k_2 k_3} \\
       0 & 0 & 0 & 0 & 0 & 0 & 0 & {k_1 k_3} &
    {k_2 k_3} & -{k_1}^2 - {k_2}^2
    \end{pmatrix}
\end{multline}
\normalsize

\section{Linearized Einstein equations in empty AdS${}_5$}
\label{A:emptyAdS5}

The Einstein equations in AdS${}_5$ can be read off of \eqref{E:EOMforX} by setting $h = 1$. In this limit, the matrix $V^T$ becomes a constant matrix. We will be interested in its five linearly independent eigenvectors
\eqn{FiveEigenvectors}{
    \chi^{1}_j & = \hbox{\tiny $\begin{pmatrix}
                \vec{k}^2-3\omega^2&0&0&0&0&\frac{1}{3}(2 k_1^2-k_2^2-k_3^2)
	&\frac{2}{3}\left(-2 k_1^2+k_2^2+k_3^2\right)
	&2\frac{k_1}{\omega} \left(k_2^2+k_3^2-2 \omega ^2\right)
	&2\frac{k_2}{\omega} \left(\omega ^2-k_1^2\right)
	&2\frac{k_3}{\omega} \left(\omega ^2-k_1^2\right)
               \end{pmatrix}$}\cr
  \chi^{2}_j&= \hbox{\tiny $\begin{pmatrix}
	0&\vec{k}^2-3\omega^2&0&0&0& k_1 k_2 &-2 k_1 k_2&\frac{k_2}{\omega}
   \left(-k_1^2+k_2^2+k_3^2-3 \omega ^2\right)&\frac{k_1}{\omega}
   \left(k_1^2-k_2^2+k_3^2-3 \omega ^2\right)&-2\frac{k_1 k_2
   k_3}{\omega }
	\end{pmatrix}$}\cr
  \chi^{3}_j&= \hbox{\tiny $\begin{pmatrix}
           0&0&\vec{k}^2-3\omega^2&0&0& k_1 k_3 &- 2 k_1 k_3 &\frac{k_3}{\omega}
   \left(-k_1^2+k_2^2+k_3^2-3 \omega ^2\right) &- 2 \frac{k_1 k_2
   k_3}{\omega }&\frac{k_1}{\omega} \left(k_1^2+k_2^2-k_3^2-3 \omega
   ^2\right)
          \end{pmatrix}$}\cr
  \chi^{4}_j&= \hbox{\tiny $\begin{pmatrix}
	0&0&0&\vec{k}^2-3\omega^2&0&-\frac{1}{3}\left(k_1^2-2 k_2^2+k_3^2\right)&\frac{2}{3}\left(k_1^2-2
   k_2^2+k_3^2\right)&2 \frac{k_1}{\omega} \left(\omega ^2-k_2^2\right) &2 \frac{k_2}{\omega} \left(k_1^2+k_3^2-2 \omega ^2\right)&2 \frac{k_3}{\omega} \left(\omega ^2-k_2^2\right)
	\end{pmatrix}$}\cr
  \chi^{5}_j&= \hbox{\tiny $\begin{pmatrix}
           0&0&0&0&\vec{k}^2-3\omega^2& k_2 k_3 &- 2 k_2 k_3 &-2\frac{
   k_1 k_2 k_3}{\omega }&\frac{k_3}{\omega} \left(k_1^2-k_2^2+k_3^2-3
   \omega ^2\right) &\frac{k_2}{\omega} \left(k_1^2+k_2^2-k_3^2-3 \omega
   ^2\right)
          \end{pmatrix}$}
}
with eigenvalues $-k^2 = \omega^2-\vec{k}^2$. The other five eigenvalues of $V^T$ are zero with geometric multiplicity 4.

By defining $Y^{i} = \chi^i_{j} X_j$, we obtain the massless scalar field equation
\begin{equation}
\label{E:EOMYi}
	\alpha^{-3}\partial_z\left(\alpha^3 \partial_z Y^{i}\right) -k^2 Y^{i} = -4 \chi^i_{j} S_j.
\end{equation}
Our method of solving \eqref{E:EOMYi} is to first solve
\begin{equation}
\label{E:Wequation}
	\alpha^{-3}\partial_z\left(\alpha^3 \partial_z W_n\right) -k^2 W_n = z^n,
\end{equation}
and then use the solutions $W_n$ to construct the solution to \eqref{E:EOMYi} through
\begin{equation}
	Y^i =  -4 \chi^i_{j} \sum_{a=-1}^{\infty} W_a S_j^{(a)}
\end{equation}
with $S_j = \sum_a S_{j}^{(a)} z^a$.\footnote{When we will discuss the empty AdS geometry as a low temperature limit of the black hole geometry then we'll need to consider only those terms in $S_j$ which are of leading order in a large momentum expansion. See section \ref{S:Short} for details.}
To solve \eqref{E:Wequation}, we first define $W_n(z) = z^2 (ik)^{-n} w_n(i k z)$, so that \eqref{E:Wequation} takes the canonical form
\begin{equation}
\label{E:wequation}
	\zeta^2 w_n^{\prime\prime}(\zeta)+\zeta w_n^{\prime} - (4-\zeta^2)w_n = \zeta^n
\end{equation}
where $\zeta \equiv i k z$. The solution to \eqref{E:wequation} with $n=3$ is $w_3(\zeta) = \frac{3\pi}{2}H_2(\zeta)$, where $H_2(\zeta)$ is the Struve function whose properties have been studied and are tabulated in the literature. See for example \cite{abramowitz+stegun}. Also note that the solution to \eqref{E:wequation} with $n=4$ is $w_4 = \zeta^2$ and the solution to \eqref{E:wequation} with $n=0$ is $w_0 = \zeta^{-2}$. By induction, we find that for $n \geq 3$ a solution to \eqref{E:wequation} is given by
\begin{subequations}
\label{E:Specialw}
\begin{align}
	w_{2m}(\zeta)&=P_{2m}(\zeta)&n&=2m\\
	 w_{2m-1}(\zeta)&=P_{2m-1}(\zeta)+(-1)^m(2m-5)!!(2m-1)!!\frac{\pi}{2}H_2(\zeta)&n&=2m-1
\end{align}
where $P_n(\zeta)$ are the polynomials
\begin{align}
	P_{2m}(\zeta)&=\sum_{k=0}^{m-2}(-1)^k 4^k\frac{m!}{(m-k)!}\frac{(m-2)!}{(m-k-2)!}\zeta^{2m-2k-2}\\
	 P_{2m-1}(\zeta)&=\sum_{k=0}^{m-3}(-1)^k\frac{(2m-1)!!}{(2m-1-2k)!!}\frac{(2m-5)!!}{(2m-5-2k)!!}\zeta^{2m-2k-3}.
\end{align}
Similarly, for $-1 \leq n \leq 2$ we find
\begin{align}
	w_{2} & = 1+4\zeta^{-2}\\
	w_{1} & = -\frac{1}{3}\zeta + \frac{\pi}{2}H_2(\zeta)\\
	w_{0} & = \zeta^{-2}\\
	w_{-1} & = -\frac{1}{3}\zeta^{-1}-\frac{1}{9}\zeta + \frac{\pi}{6}H_2(\zeta).
\end{align}
\end{subequations}

Equations \eqref{E:Specialw} are special solutions to the non homogeneous equation \eqref{E:wequation}. The most general solution to \eqref{E:wequation} are those given in \eqref{E:Specialw} in addition to the solutions to the homogeneous equations $J_2(\zeta)$ and $Y_2(\zeta)$ which are Bessel functions of order 2. So we should take
\begin{equation}
\label{E:wtowBessel}
 	w_n \to w_n + a_n J_2(\zeta) + b_n Y_2(\zeta).
\end{equation}
Since $Y_2 = -\frac{4}{\pi}\zeta^{-2}+\ldots$, then the boundary condition $H_{mn}(0)=0$, which translates to $\lim_{\zeta\to 0}\zeta^2 \omega_n(\zeta) = 0$, implies that
\begin{align}
\label{E:bvals}
	b_{0} &= \frac{1}{4}\pi\\
	b_{2} &= \pi
\end{align}
and $b_n=0$ for all other $n$. The coefficients $a_n$ can be found by considering spacelike momenta, $k^2>0$. 
Then,
\begin{equation}
 	H_2(i k z) + a J_2(i k z) = -i L_2(k z) - a I_2(k z)
\end{equation}
where $L_2$ is a modified Struve function and $I_2$ is a modified Bessel function. Both $L_2$ and $I_2$ diverge exponentially as we take $z \to \infty$, but the combination $L_2 - I_2$ remains finite. Thus,
\begin{equation}
\label{E:avals}
	a_{n} = \begin{cases}
          - i & n = 2 m -1\\
	  0 & n = 2m.
        \end{cases}
\end{equation}
Now that the solution to \eqref{E:wequation} is available, we can easily solve \eqref{E:EOMYi},
\begin{equation}
	Y^i = -4 \chi^i_{j} \sum_{a=-1}^{\infty} z^2 w_{a}(z q) S_j^{(a)} (ik)^{-a}.
\end{equation}

For completeness, we note that the remaining five equations of motion will leave us with five undetermined integration constants which, as described in section \ref{SS:BoundaryStress}, are determined by the five first order constraint equations.

\section{Holographic renormalization}
\label{A:HoloRG}
The prescription for obtaining the boundary theory stress tensor $\langle T_{mn} \rangle$ from an asymptotically AdS bulk metric has been developed in \cite{Balasubramanian:1999re,deHaro:2000xn} following the basic prescription of \cite{Gubser:1998bc,Witten:1998qj} and is given by
\begin{equation}
\label{E:onepoint}
 	\langle T_{mn} \rangle = \lim_{\epsilon \to 0}\frac{2}{\sqrt{g(\vec{x},\epsilon)}}\frac{\partial S_{ren}}{\partial g^{mn}(\vec{x},\epsilon)}
\end{equation}
where $g^{mn}(\vec{x},z)$ is defined through the line element
\begin{equation}
\label{E:GrFe}
 	ds^2 = \frac{L^2}{z^2}\left(dz^2+g_{mn}(\vec{x},z)dx^m dx^n\right)
\end{equation}
and $z \to 0$ corresponds to the asymptotic boundary. We take $\epsilon$ to be small so that
\begin{equation}
    \gamma_{mn}(\vec{x},\epsilon)=\frac{L^2}{\epsilon^2}g_{mn}(\vec{x},\epsilon)
\end{equation}
gives us the metric on a spacelike hypersurface of constant $z=\epsilon$ close to the conformal boundary (and it is understood that $\gamma_{5\mu}=0$.) The action $S_{ren}$ is given by
\begin{equation}
    S_{ren}= S_{EH}+S_M+S_{GH}+S_{ct}.
\end{equation}
The Einstein-Hilbert action $S_{EH}$ and the matter action $S_M$ are given in \eqref{GenericAction}. The Gibbons-Hawking boundary term reads
\begin{equation}
	S_{GH} = -\frac{1}{\kappa_5^2}\int_{z=\epsilon} d^4x \sqrt{\gamma} K
\end{equation}
with $K$ the extrinsic curvature,
\begin{align}
\label{E:Extrinsic}
    K_{\mu\nu}&\equiv\gamma_{\mu}^{\sigma}D_{\sigma}n_{\nu}=-\frac{\epsilon}{2L}\partial_\epsilon \gamma_{\mu\nu}
\intertext{and}
    n^{\mu}&=-\frac{L}{\epsilon}\delta^{\mu}_5.
\end{align}
The boundary action, $S_{ct}$, is constructed from the boundary values of the dynamical fields. In the probe approximation which we are working in, the only dynamical field is the boundary metric $\gamma_{mn}$ so that $S_{ct}=S_{ct}[\gamma]$.
More explicitly,
\begin{equation}
\label{E:SctExplicit}
    S_{ct} = -\frac{1}{2\kappa_5^2}\int_{z=\epsilon} d^4x \sqrt{\gamma}\left(\frac{2}{L} {\rm A}_{\Lambda}+R^{\gamma}L {\rm A}_{R}\right)
\end{equation}
with ${\rm A}_{\Lambda}$ and ${\rm A}_{R}$ undetermined numbers and $R^{\gamma}$ is the Ricci Scalar associated with the boundary metric $\gamma_{mn}$. It is easy to convince oneself that, in the setup we're considering, any higher order terms in derivatives of $\gamma$ will not contribute to $\langle T_{mn}\rangle$. If a conformal anomaly is present then $S_{ct}$ can also depend explicitly on $\ln\epsilon$ \cite{deHaro:2000xn}. Usually a logarithmic correction to $\gamma_{mn}$ at order $\mathcal{O}(\epsilon^2)$ is indicative of such an anomaly. In our case, if $J_{mn}^{(0)} \neq 0$ or $J_{mn}^{(2)} \neq 0$ such a logarithmic term will appear in the series expansion for $H_{mn}$ and we will find a trace anomaly for the boundary stress tensor (see \eqref{E:Tmn}.) However, since in our case the logarithmic terms are not induced by a deformation of the boundary theory, as in \cite{deHaro:2000xn}, we will need to deal with the $\ln\epsilon$ divergences (and other divergences) in a somewhat different manner.

\subsection{Evaluating $\langle T_{mn}^{GH} \rangle$}
The program of holographic renormalization \cite{Bianchi:2001de,Bianchi:2001kw} provides a prescription for choosing $S_{ct}$ such that all the correlation functions of the boundary theory will be finite. In the current context, this amounts to finding the two coefficients ${\rm A}_{\Lambda}$ and ${\rm A}_{R}$ in \eqref{E:SctExplicit}. Thus, it is much simpler to find the appropriate counterterm action $S_{ct}$ by requiring that the one-point function \eqref{E:onepoint} is finite in the $\epsilon \to 0$ limit. This method has been used in \cite{Aharony:2005zr} to find the counterterm action in the presence of irrelevant operators. Here, the benefit is that we do not have to deal with the unknown contribution of $S_M$ to the action when evaluated on-shell:
\begin{align}
    \langle T_{mn}^{GH} \rangle &=
        \lim_{\epsilon \to 0}\frac{2}{\sqrt{g(\vec{x},\epsilon)}}\frac{\partial }{\partial g^{mn}(\vec{x},\epsilon)}\left(S_{EH}+S_M+S_{GH}\right)\\
        &=
        \lim_{\epsilon \to 0}\frac{2}{\sqrt{g(\vec{x},\epsilon)}}\frac{\partial }{\partial g^{mn}(\vec{x},\epsilon)} S_{GH}\\
\label{E:TijGH}
        &=
        -\lim_{\epsilon \to 0}\frac{L^2}{\kappa_5^2\epsilon^2}\left(K_{mn}-K\gamma_{mn}\right).
\end{align}
Using \eqref{E:Extrinsic} we find
\begin{equation}
\label{E:TGH_2}
	\langle T_{mn}^{GH}\rangle
	=
	-\lim_{\epsilon \to 0}\frac{L^3}{\kappa_5^2\epsilon^2}
	\left(-\frac{3}{\epsilon^2}g_{mn}+\frac{g_{mn} g^{ts}\partial_{\epsilon}\,g_{ts}-\partial_{\epsilon}\,g_{mn}}{2\epsilon}
	\right).
\end{equation}
Rewriting the metric \eqref{E:Gmn} in the Graham-Fefferman coordinate system \eqref{E:GrFe},
\begin{equation}
\label{E:gToH}
	g_{mn}=\eta_{mn}+ \frac{\kappa_5^2}{2 L^3} \langle T_{mn}\rangle_{\rm bath} \epsilon^4 + \frac{\kappa_5^2}{2 L^3}H_{mn}+\mathcal{O}(\epsilon^5)\,,
\end{equation}
\eqref{E:TGH_2} reduces to
\begin{equation}
\label{E:TGH_3}
\langle T_{mn}^{GH}\rangle
	=
	\lim_{\epsilon \to 0}
	\left(
	\frac{L^3}{\kappa_5^2}\frac{3 g_{mn}}{\epsilon^4}+\frac{1}{4\epsilon^{3}}\partial_{\epsilon}\left(H_{mn}-\eta_{mn}H^{l}_{l}\right)\right)+\left(\substack{\hbox{ finite}\\\hbox{terms}}\right).
\end{equation}
In what follows we expand $H_{mn}$ in a power series, as in \eqref{E:HSeries}, but with an additional $H_{mn}(0)=H_{mn}^{(0)}\neq 0$ term which we shall set to zero at the end of the day. The reason we temporarily allow such a term will become clear shortly.
To express \eqref{E:TGH_3} in terms of the energy momentum tensor of the probe source, we use \eqref{E:HmnToJ} to substitute the $H_{mn}^{(a)}$'s with the $J_{mn}^{(a)}$'s.\footnote{The astute reader might worry that equation \eqref{E:HmnToJ} has been constructed for the $H_{mn}^{(0)}=0$ case. As it is written, it is also valid for non vanishing $H_{mn}^{(0)}$ (and thus non vanishing $R_{mn}^{(0)}$) as long as we are working to linear order in $H_{mn}$.} Working withe boundary metric $\gamma_{mn}=\frac{L^2}{\epsilon^2}g_{mn}$ instead of $g_{mn}$,
we find that the divergent terms in $\langle T_{mn}^{GH}\rangle$ are given by
{\small
\begin{multline}
\label{E:TGH_4}
	-\frac{L^2}{\kappa_5^2\epsilon^2}\left(
		 -\frac{3}{L}\gamma_{mn}+\frac{L}{2}\left(R^{\gamma}_{mn}-\frac{1}{2}R^{\gamma}\gamma_{mn}\right)
		\right)
	+\frac{J_{mn}^{(-1)}}{3 \epsilon^3}+\frac{J_{mn}^{(0)}}{2 \epsilon^2}+\frac{J_{mn}^{(1)}}{\epsilon}-J_{mn}^{(2)}\ln\epsilon/L
	\\
	-\frac{L^2}{\kappa_5^2\epsilon^2}\left(	
		 \frac{L}{2}\left(R^{\gamma\,(1)}_{mn}-\frac{1}{2}R^{\gamma\,(1)}\gamma_{mn}\right)\epsilon
		 -L\left(R^{\gamma\,(2)}_{mn}-\frac{1}{2}R^{\gamma\,(2)}\gamma_{mn}\right)\epsilon^2\ln\epsilon/L
		\right)
\end{multline}
}where $R^{\gamma}=\frac{\epsilon^2}{L^2}R$ and $R^{\gamma}_{mn} = R_{mn}$ are the Ricci scalar and Ricci tensor corresponding to the boundary metric $\gamma_{mn}$.

\subsection{Evaluating the counterterm action}
Some of the divergences in \eqref{E:TGH_4} can be removed by an appropriate choice of ${\rm A}_{\Lambda}$ and ${\rm A}_{R}$ in \eqref{E:SctExplicit}. Comparing
\begin{equation}
       \lim_{\epsilon \to 0}\frac{2}{\sqrt{g(\vec{x},\epsilon)}}\frac{\partial }{\partial g^{mn}(\vec{x},\epsilon)}S_{ct}=
	-\frac{L^2}{\kappa_5^2\epsilon^2}\left(-\frac{{\rm A}_{\Lambda}}{L}+L {\rm A}_{R} \left(R^{\gamma}_{ij}-\frac{1}{2}R^{\gamma} \gamma_{ij}\right)\right)
\end{equation}
to \eqref{E:TGH_4}, we find that to get rid of the leading divergences we must set
\begin{equation}
\label{E:ctterms1}
 	{\rm A}_{\Lambda}=-3,\quad {\rm A}_{R}=-\frac{1}{2}.
\end{equation}
Had we've taken $H_{mn}^{(0)}=0$, we would have concluded that
\begin{equation}
\label{E:ctterms2}
 	{\rm A}_{\Lambda}=-3,\quad {\rm A}_{R}=-1;
\end{equation}
once $H_{mn}(0) = 0$ the leading divergences coming from the $\left(R^{\gamma}_{mn}-\frac{1}{2}R^{\gamma}\gamma_{mn}\right)$ terms are $L\left(R^{\gamma\,(1)}_{mn}-\frac{1}{2}R^{\gamma\,(1)}\gamma_{mn}\right)\epsilon^{-1}$. Setting $H_{mn}^{(0)} \neq 0$ implies that the leading divergent terms are $\frac{L}{2}\left(R^{\gamma\,(1)}_{mn}-\frac{1}{2}R^{\gamma\,(1)}\gamma_{mn}\right)\epsilon^{-2}$.
Since the counterterm action should be independent of the boundary metric, we should use the same $S_{ct}$ no matter what the boundary conditions on $H_{mn}(0)$ are (as long as they're finite.). Also,
by setting $J_{mn}=0$ we should get the counterterms needed for a pure gravity theory with a non flat boundary metric. These have been worked out in \cite{deHaro:2000xn} and coincide with \eqref{E:ctterms1} once we take into account the appropriate conventions.
Thus, we must use \eqref{E:ctterms1}.

At this point we are still left with a stress-energy tensor which includes power law and logarithmic divergences. In the pure gravity theory logarithmic divergences may be canceled by a boundary action containing four derivatives of the boundary metric $\gamma$. The reason these kinds of counterterms are available in the pure gravity case is that there, $\left(R^{\gamma}_{mn} - \frac{1}{2}R^{\gamma} \gamma_{mn}\right)^{(2)}$ may be written as a certain combination of $R^{\gamma\,(0)}_{mn}$, $R^{\gamma\,(0)}$ and $\gamma^{(0)}_{mn}$ and this combination may be obtained by varying a higher derivative boundary action \cite{deHaro:2000xn}. While we may add such higher derivative terms to \eqref{E:SctExplicit}, they will not cancel any of the remaining divergences in \eqref{E:TGH_4} since the latter are independent of $R^{\gamma\,(0)}_{mn}$, $R^{\gamma\,(0)}$ and $\gamma^{(0)}_{mn}$.
Put differently, the probe-source acts as an external current in the equations of motion. As a result, the divergences it introduces can not be subtracted away by counterterms constructed only of the metric tensor. Luckily, since the $J_{mn}^{(a)}$'s have finite spatial extent, these divergences can be thought of as divergent contact terms, and we expect that they are associated with formally divergent parameters of the source.

After setting $H_{mn}(0)=0$ and separating the divergent terms $\langle T_{mn}^{\epsilon}\rangle$ from the finite terms of the stress tensor in \eqref{E:onepoint}, we find 
\begin{equation}
	\langle T_{mn} \rangle
	=
	\langle T_{mn}\rangle_{\rm bath}
	+\langle T_{mn}^{\epsilon} \rangle
	+H_{mn}^{(4)} - \eta_{mn}H^{(4)\,s}_s + \frac{3 L^3}{4\kappa_5^2}\left(R_{mn}^{(2)}-\frac{1}{2}R^{(2)}\eta_{mn}\right)-\frac{1}{4}J_{mn}^{(2)}
\end{equation}
with
\begin{multline}
\label{E:Tmnepsilon}
	\langle T_{mn}^{\epsilon} \rangle =
		\frac{J_{mn}^{(-1)}}{3 \epsilon^3}+\frac{J_{mn}^{(0)}}{2 \epsilon^2}+\frac{J_{mn}^{(1)}}{\epsilon}-J_{mn}^{(2)}\ln\epsilon/L
\\
	-\frac{L^3}{\kappa_5^2}\left(	
		 \frac{1}{2}\left(R_{mn}^{(1)}-\frac{1}{2}R^{(1)}\eta_{mn}\right)\frac{1}{\epsilon}
		 -\left(R_{mn}^{(2)}-\frac{1}{2}R^{(2)}\eta_{mn}\right)\ln\epsilon/L
		\right).
\end{multline}

\end{appendix}

\clearpage
\bibliographystyle{ssg}
\bibliography{source}

\end{document}